\documentclass[aps,pra,superscriptaddress,twocolumn]{revtex4}

\usepackage{amstext,amsmath,amssymb,ulem}
\usepackage{euscript}
\usepackage{color}
\usepackage{float}
\restylefloat{table}

\usepackage[T1]{fontenc}
\usepackage[utf8]{inputenc}
\usepackage{graphicx, subfigure}
\usepackage{mathptm}
\usepackage[colorlinks]{hyperref}

\newcommand{\beq}{\begin{equation}}
\newcommand{\eeq}{\end{equation}}
\newcommand{\ket} [1] {\vert #1 \rangle}
\newcommand{\bra} [1] {\langle #1 \vert}

\newcommand{\tr}{\mathop{\mathrm{tr}}}

\newcommand{\ba}{\begin{align}}
\newcommand{\ea}{\end{align}}
\newcommand{\bea}{\begin{eqnarray}}
\newcommand{\eea}{\end{eqnarray}}

\newcommand{\scale}{[\vec{s}]n}
\newcommand{\scalezero}{[\vec{s}]0}

\newcommand{\waver}[1]{[\vec{w}]#1}

\makeatletter
\@ifundefined{textcolor}{}
{%
 \definecolor{BLACK}{gray}{0}
 \definecolor{WHITE}{gray}{1}
 \definecolor{RED}{rgb}{1,0,0}
 \definecolor{GREEN}{rgb}{0,.4,0}
 \definecolor{BLUE}{rgb}{0,0,1}
 \definecolor{CYAN}{cmyk}{1,0,0,0}
 \definecolor{MAGENTA}{cmyk}{0,1,0,0}
 \definecolor{YELLOW}{cmyk}{0,0,1,0}
 }

\makeatother

\usepackage{bm}
\usepackage{mathtools}

\def\id{I}

\def\1{\mat{\id}}

\def\mat#1{\mathbf{#1}}

\renewcommand{\vec}[1]{\bm{\mathrm{#1}}}

\renewcommand{\sout}[1]{}

\begin{document} 
\title{Holographic Construction of Quantum Field Theory using Wavelets}
\author{Sukhwinder Singh and Gavin K. Brennen}
\affiliation{Centre for Engineered Quantum Systems, Department of Physics and Astronomy, Macquarie University, North Ryde, NSW 2109, Australia}

\begin{abstract}
Wavelets encode data at multiple resolutions, which in a wavelet description of a quantum field theory, allows for fields to carry, in addition to space-time co\"ordinates, an extra dimension: scale. A recently introduced Exact Holographic Mapping [C.H. Lee and X.-L. Qi, Phys. Rev. B 93, 035112 (2016)] uses the Haar wavelet basis to represent the free Dirac fermionic quantum field theory (QFT) at multiple renormalization scales thereby inducing an emergent bulk geometry in one higher dimension. This construction is, in fact, generic and we show how higher families of Daubechies wavelet transforms of $1+1$ dimensional scalar bosonic QFT generate a bulk description with a variable rate of renormalization flow. In the massless case, where the boundary is described by conformal field theory, the bulk correlations decay with distance consistent with an Anti-de-Sitter space (AdS$_{3}$) metric whose radius of curvature depends on the wavelet family used. We propose an experimental demonstration of the bulk/boundary correspondence via a digital quantum simulation using Gaussian operations on a set of quantum harmonic oscillator modes. 
\end{abstract}


\maketitle

\section{Introduction}
\label{intro}
The holographic principle asserts that spacetime may be like a hologram: information contained within the volume of spacetime can be encoded on the boundary of that spacetime \cite{'tHooft,Susskind,Bouso:02}. This heuristic principle has been used with much success in the study of black hole thermodynamics and quantum gravity, and has been precisely formulated as holographic dualities, in particular, the Anti-de-Sitter space/conformal field theory (AdS/CFT) correspondence \cite{adscft}. 

A holographic duality (e.g., the AdS/CFT correspondence) is a duality or equivalence between a theory with gravity and a theory without gravity that are seen to live in the bulk and at the boundary of a spacetime manifold respectively. The extra dimension (that generates the bulk) corresponds to energy scale of the boundary theory, and  the gravitational dynamics in the bulk generalize the renormalization group flow equations of the boundary. A characteristic prediction of the holographic principle is the existence of correlated noise in the measured positions of massive bodies, though experimental evidence so far is negative \cite{Chou:15}. 

Recently, Qi \emph{et al.} proposed a concrete implementation of such a bulk/boundary correspondence. Their proposal, called the \textit{Exact Holographic Mapping} (EHM) \cite{Qi:13, QiLee:16}, maps the (boundary) theory of a Dirac fermion descretized on a lattice to a bulk theory with the same number of degrees of freedom (DOFs) as the boundary, but using wavelet and scale field mode operators which carry a spatial position index as well as a scale index. They used Haar wavelets, which is the $\mathcal{K}=1$ family of Daubechies wavelets \cite{Daubechies}. This EHM of the fermionic theory demonstrates some key features of holographic duality such as an emergent geometry in bulk---as inferred from the scaling of correlations between bulk wavelet DOFs---that depends on the boundary theory. In this way, the authors were able to demonstrate the emergence of a bulk AdS geometry from a CFT on the boundary, the emergence of a flat metric from a massive boundary quantum field theory (QFT), and some features of black hole physics in the bulk for the case of a thermal CFT at the boundary. 

In this work, we show that the EHM extends to an entire class of wavelet transformations. We do this by analyzing the Daubechies $\mathcal{K}$ wavelet families for $\mathcal{K}\geq 3$. Here the $\mathcal{K}$ index refers to the number of vanishing moments of the constituent scale and wavelet functions. A larger index encodes a function more accurately at the cost of increased computational effort. We show how this feature translates to changing the scaling of correlations in the bulk description of a boundary QFT such that larger $\mathcal{K}$ provides faster renormalization of the theory, at the expense of a linear increase in the interaction neighborhood for each renormalization step. Furthermore, using $\mathcal{K}\geq 3$ ensures that the derivative couplings can be captured exactly up to the cutoff scale since the wavelet and scale functions then have continuous first derivatives.

We conclude the introduction by remarking that the wavelet approach bears interesting connections with implementations\cite{Swingle,Qi:13} of the holographic principle in the tensor network formalism. As noted in Ref.~\cite{Qi:13}, the Haar wavelet transform can be described as a \textit{tree tensor network}---a set of tensors (multi-linear transformations) that are contracted according to a tree graph. In this description, a tensor at a given renormalization layer (radius) takes as input two nearest neighbor fields and outputs a short range wavelet field and a longer range (coarse grained) scale field used as input to the next stage of renormalization. In this work, we generalize to Daubechies $\mathcal{K}$ wavelet families, which go beyond tree tensor networks and are closer in spirit to tensor networks with loops such as the \textit{Multi-Scale Renomalization Ansatz} (MERA) \cite{Guifre:07,cMERA} (of which, a tree tensor network is a special case corresponding to fixing some of the MERA tensors to the identity). In fact, in Ref.~\cite{GlenWhite}, the authors establish a precise way to describe various wavelet transforms as MERA tensor networks. 

The plan of the paper is as follows. We begin with a brief introduction to the basics of the wavelets functional basis in Sec. \ref{waveletbasics} and why it is a natural one for holographic encoding. In Sec. \ref{bosonicQFT} we describe the bulk and boundary Hamiltonians for free scalar bosonic QFT. Sec. \ref{BB} presents the main results given the details of the bulk/boundary correspondence and the shape of the emergent bulk geometry. Explicit circuit based constructions of the bulk and boundary states are provided in Sec. \ref{Circuit}. The requisite components are local ground or thermal state preparation of an array of bosons, linear optical gates, and single mode squeezing which could be done in a variety of engineered atomic, optical, or solid state platforms. Finally, we conclude with a summary and comment on open problems.

\section{The wavelet basis}
\label{waveletbasics}

\begin{figure}[t]
	\begin{center}
		\includegraphics[width=\columnwidth]{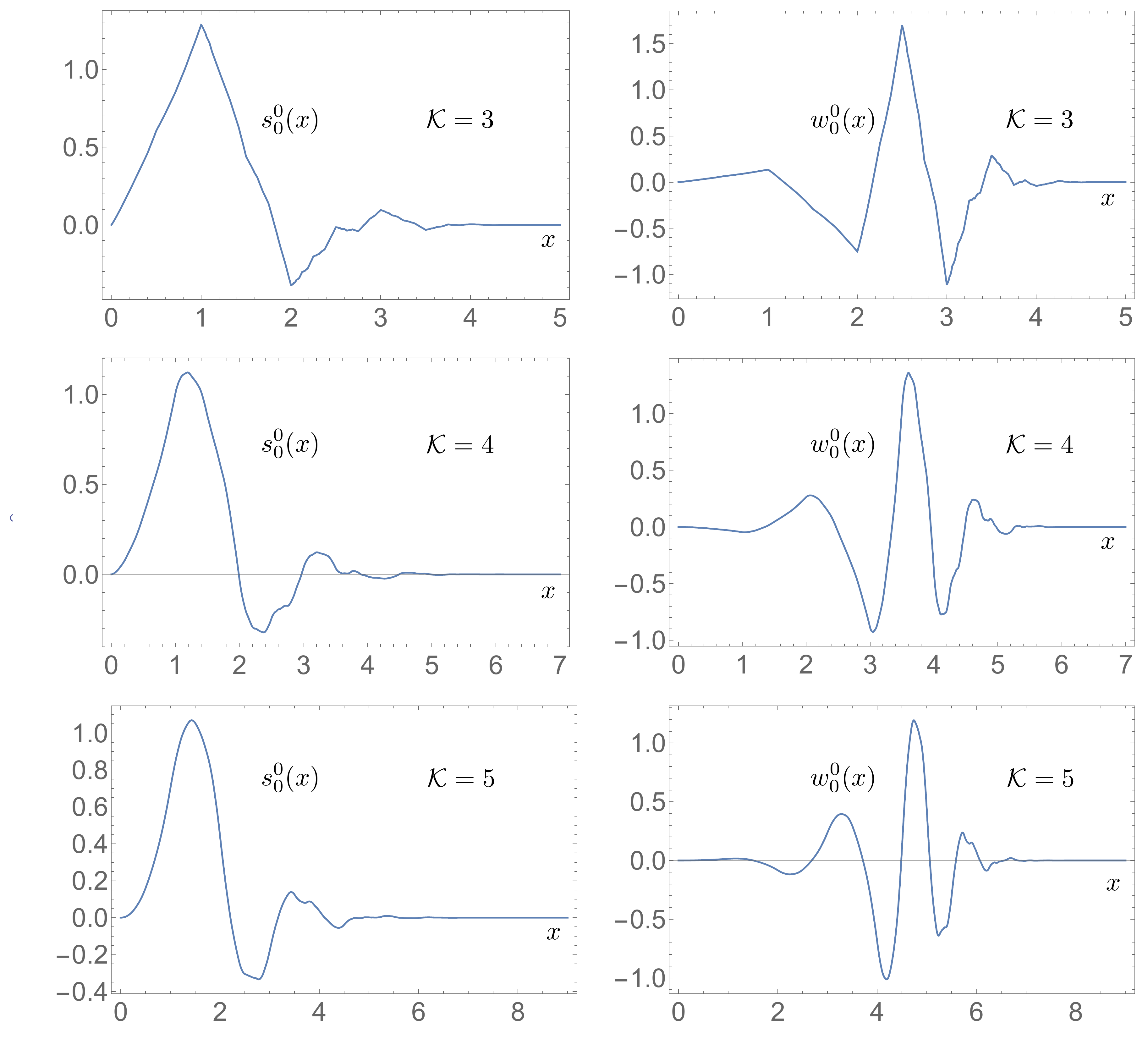}
	\end{center}
	\caption{Plots of scale functions $s(x)$ and wavelet functions $w(x)$ for Daubechies families $\mathcal{K}=3,4,5$.}
	\label{fig:WaveletPlots}
\end{figure}

Wavelets constitute an orthonomal basis for the Hilbert space $L^2(\mathbb{R})$, square integrable functions on the line, and we briefly review some of their properties here. For a comprehensive survey see Ref.~\cite{Mallat}.  Generically, wavelets  are defined in terms of a mother wavelet function $w(x)$ and a father scaling function $s(x)$ by taking linear combinations of shifts and rescalings thereof. For the remainder we focus on one family known as Daubechies $\mathcal{K}$-wavelets where the role of the integer $\mathcal{K}$ will be described below.  First we introduce two unitary operators on $L^2(\mathbb{R})$:  $\mathcal{T}$ for  discrete translation and $\mathcal{D}$ for scaling defined by the action on a function $f\in L^2(\mathbb{R})$:
\begin{equation}
\mathcal{D}f(x)=\sqrt{2}f(2x);\quad \mathcal{T}f(x)=f(x-1).
\end{equation}
The father scaling function $s(x)$ is a solution to the linear renormalisation group equation
\begin{equation}\label{Father}
s(x)=\mathcal{D}\left[\sum_{n=0}^{2\mathcal{K}-1}h_n\mathcal{T}^n s(x)\right],
\end{equation}
which performs block averaging followed by rescaling. The $2\mathcal{K}$ real coefficients $\{h_n\}$ are computed analytically for $\mathcal{K}< 4$ and are solved for numerically otherwise. For convenience, we choose language such that $\emph{scale}$ refers to a number $r\in \mathbb{N}$ which describes features at a resolution $2^{-r}$ in some natural units, and higher(lower) scale means larger(smaller) $r$.
Given the solution to $s(x)$, higher scaling functions are defined by applying $n$ unit translations followed by $k$ scaling transformations on the father:
\begin{equation}
s^k_n(x) =\mathcal{D}^k \mathcal{T}^n s(x).
\end{equation}
The scaling functions are normalised so that
\begin{equation}
\int dx\ s^k_n(x)=1.
\end{equation}
The mother wavelet $w(x)$ and the father $s(x)$ have the property that they are neither localised in position or momentum.   The wavelets take the following form:
\begin{equation}
w(x)=\sum_{n=0}^{2\mathcal{K}-1}g_n\mathcal{D}\mathcal{T}^n s(x)=\sum_{n=0}^{2\mathcal{K}-1}g_ns^1_n(x),
\end{equation}
where the set of coefficients $\{g_n\}$ are obtained from $\{h_n\}$ by reversing the order and alternating signs:  $g_n=(-1)^nh_{2\mathcal{K}-1-n}$ (see Appendix \ref{waveletprops}).  Scale $k$ wavelets are obtained by translating and scaling the mother:
\begin{equation}
w^k_n(x)=\mathcal{D}^k \mathcal{T}^n w(x).
\end{equation}
The index $\mathcal{K}$ specifies the number of vanishing moments of the wavelets, i.e. 
\[
\int\ dx\ w(x)x^p=0\quad p=0,..,\mathcal{K}.
\]
The vanishing of the zeroth moment guarantees that the wavelet basis is square integrable~\cite{Mallat}.
Choosing larger $\mathcal{K}$ means more features can be captured at a given scale, however at the expense of additional computational cost since more translations are needed during block averaging. Daubechies wavelets are optimal in the sense that they have the smallest size support for a given number of vanishing moments~\cite{Mallat}. 
The basis functions $s^k_n(x)$ and $w^k_n(x)$ have support on $[2^{-k}n,2^{-k}(n+2\mathcal{K}-1)]$ and satisfy the following orthonormality relations:
\begin{equation}
\begin{split}
&\int dx\ s^k_n(x)s^k_m(x)=\delta_{m,n},\\
&\int dx\ s^k_n(x)w^{k+l}_m(x)=0\quad (l\geq 0),\\
&\int dx\ w^k_n(x)w^l_m(x)=\delta_{m,n}\delta_{k,l}.
\end{split}
\label{orthonormal}
\end{equation}
By the last relation, the wavelets constitute normalised wave functions.  The scaling functions at scale $2^{-k}$ are complete in that
\begin{equation}
\sum_{n=-\infty}^{\infty} \frac{1}{\sqrt{2^k}}s^k_n(x)=1.
\end{equation}
A final important property of the Daubechies $\mathcal{K}$-wavelets is that they are $\mathcal{K}-2$ times differentiable. A plot of three of the the scale and wavelet functions families is shown in Fig. \ref{fig:WaveletPlots}.

The connection between holography and wavelets arises from the scale dependent representation of fields. 
In a wavelet family basis, linear superpositions of scale functions $\{s^r_m(x)\}_{m=-\infty}^{\infty}$ (with square summable coefficients) span a subspace $\mathcal{H}_r$ of $L^2(\mathbb{R})$.
This is a proper subspace of the higher scale space $\mathcal{H}_r\subset \mathcal{H}_{r+j}\ (j>0)$.  Linear combinations of the scale $r$ wavelet functions $\{w^r_m(x)\}_{m=-\infty}^{\infty}$ span the orthocomplement, $\mathcal{W}_r$, of $\mathcal{H}_r$ in $\mathcal{H}_{r+1}$:
$\mathcal{H}_{r+1}=\mathcal{H}_{r}\oplus\mathcal{W}_{r}$.  We can use a set of scaling functions $\{s^r_m(x)\}_{m=-\infty}^{\infty}$ to represent features at scale $r$ and a set of wavelets $\{w^r_m(x)\}_{m=-\infty}^{\infty}$ to represent features at scale $r+1$ that cannot be represented at scale $r$.
The whole space has the following decomposition satisfied for \emph{any} finite $r$:
\begin{equation}
L^2(\mathbb{R})=\mathcal{H}_{r}\bigoplus_{l=r}^{\infty}\mathcal{W}_{l},
\label{Hilbertspace}
\end{equation}
meaning that for a fixed scale $r$ the set 
\[
\{s^r_m(x)\}_{m=-\infty}^{\infty}\bigcup \{w^l_m(x)\}_{m=-\infty,l=r}^{\infty,\infty},
\]
 span a basis for $L^2(\mathbb{R})$ \cite{Mallat}.
 
 In particular, if one introduces a scale cutoff $n$, then 
 \begin{equation}
 \mathcal{H}_0\oplus_{l=0}^{n-1} \mathcal{W}_l\simeq  \mathcal{H}_n.
 \end{equation}
What this means is that the Hilbert space of functions spanned by scale fields at scale $n$ is isomorphic to the Hilbert space spanned by wavelet functions at all lower scales together with scale functions at the coarsest scale. We will identify the Hilbert space $\mathcal{H}_n$ as characterizing the Hilbert space of quantum fields with a UV cutoff momentum $p\simeq 2^n/L$, where $L$ is the length of the system in units of distance at the coarsest scale. We will call the degrees of freedom (DOFs) in this Hilbert space to be \emph{boundary} DOFs. Similarly, the DOFs at lower scale in $\mathcal{H}_0\oplus_{l=0}^{n-1} \mathcal{W}_l$ are \emph{bulk} DOFs. Since the Hilbert spaces are isomorphic, there exists an unitary transformation from boundary states to bulk states. 

\begin{figure}[t]
	\begin{center}
		\includegraphics[width=\columnwidth]{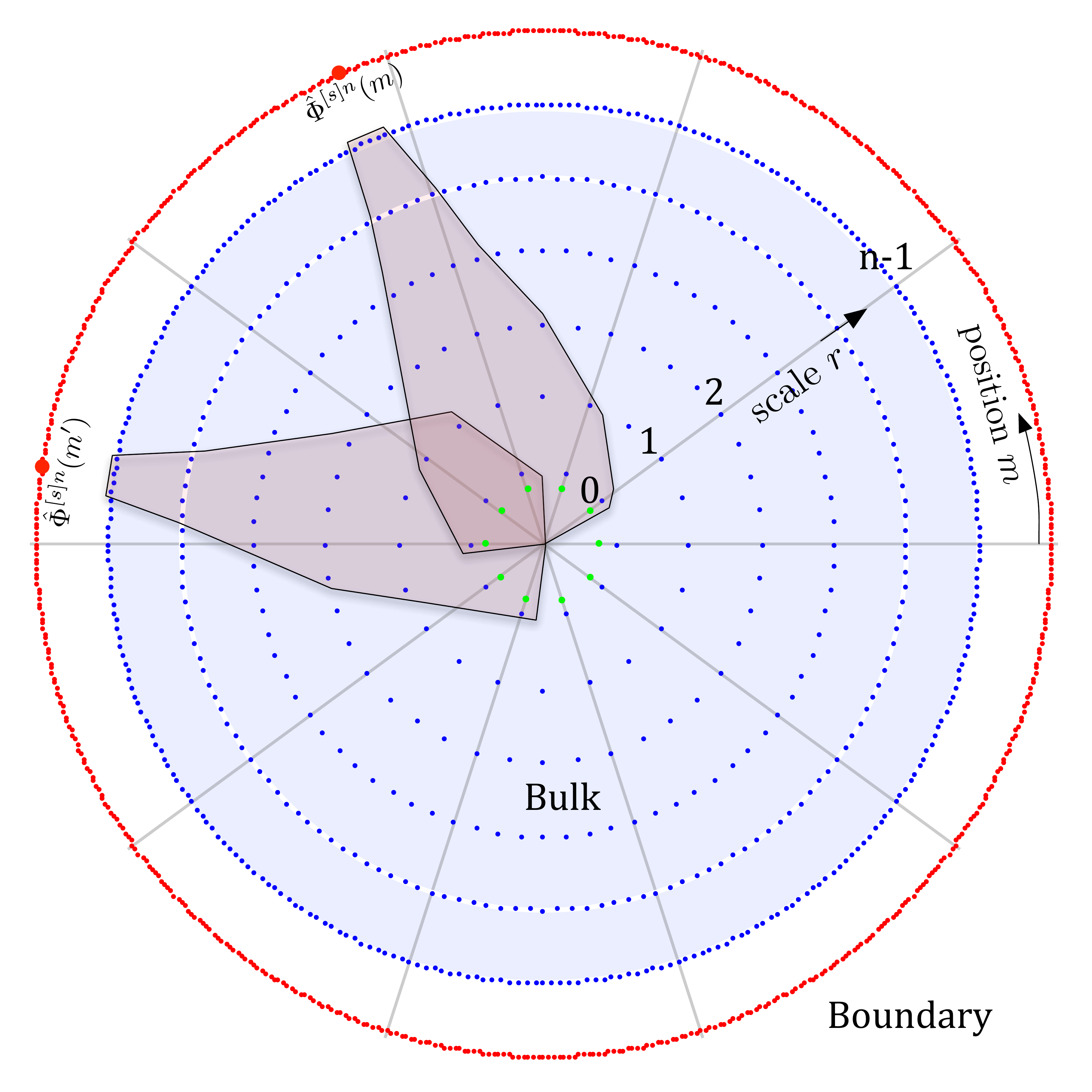}
		\label{fig:masslessopenent}
	\end{center}
	\caption{Bulk/boundary corrrepondence for a time slice of a 1+1 dimensional QFT with periodic boundaries. The QFT has a UV cutoff at scale $n$, and the boundary state $\ket{G}_{\rm bd}$ is the ground state of a boundary Hamiltonian $\hat{H}_{\rm bd}$ with DOFs $\{\hat{\Phi}^{[s]n}(\theta)\}$ (red dots) . The corresponding bulk state $\ket{G}_{\rm bk}$ is the ground state of the Hamiltonian $\hat{H}_{\rm bk}$ with DOFs $\{\hat{\Phi}^{[s]0}(m)\}\cup \{\hat{\Phi}^{[w]r}(m)\}$ (green and blue dots). The number $L$ of DOFs at the coarsest scale $r=0$ in the bulk is determined by the wavelet family used. For the field $\hat{\Phi}^{[w]r}(m)$ at scale $r$ and position $m$, one can associate the spatial AdS$_{3}$ co\"ordinates $(\rho=\frac{L2^{r}}{2\pi},\theta=\frac{m2\pi}{L2^{r}})$. The shaded segments within the blue region enclose the bulk DOFs that contribute to the two point correlation $_{\rm bd}\bra{G}\hat{\Phi}^{\scale}(m)\hat{\Phi}^{\scale}(m')\ket{G}_{\rm bd}$. The boundary fields overlap with the same number ($2\mathcal{K}-1$) of bulk fields at each scale, here it is $5$ for the $\mathcal{K}=3$ wavelet family.}
\end{figure}

\section{1+1 D Scalar Bosonic QFT}
\label{bosonicQFT}
To illustrate the connection between holography and wavelets we study scalar bosonic free field theory in one spatial dimension. The Hamiltonian for this QFT is:
\begin{equation}
\hat{H}=\int dx\ \frac{1}{2}\left(\hat{\Pi}^2(x,t)+\vec{\nabla}\hat{\Phi}^2(x,t)+m_0^2\hat{\Phi}^2(x,t)\right),
\label{Ham}
\end{equation}
and the canonical momentum is 
\begin{equation}
\hat{\Pi}(x,t)=\frac{\partial \hat{\Phi}(x,t)}{\partial t},
\end{equation}
which together with the field $\hat{\Phi}(x,t)$ is normalised to satisfy the equal time commutation relation $[\hat{\Phi}(x,t),\hat{\Pi}(y,t)]=i\delta(x-y)$ $(\hbar\equiv 1)$.
Here the phase velocity of waves in this theory is set so that the speed of light is $1$ and the bare mass is $m_0$.

We will focus on a system of size $V$ with periodic boundaries such that $\hat{\Phi}(x+V,t)=\hat{\Phi}(x,t)$. This choice obviates the need to define special boundary wavelets and allows us to make a connection to geometry in the bulk even for systems with finite numbers of DOFs. The Hamiltonian $\hat{H}$ is diagonalized with the normal mode operators $\hat{a}_{k_n}$ and $\hat{a}^{\dagger}_{k_n}$ satisfying $[\hat{a}_{k_n},\hat{a}^{\dagger}_{k_m}]=\delta_{n,m}$, where $k_n=\frac{2\pi n}{V}$ with $n\in\mathbb{Z}\cap[-V/2,V/2]$. The fields operators are then (henceforth we suppress the time dependence)
\[
\begin{array}{lll}
\hat{\Phi}(x)&=&\frac{1}{\sqrt{V}}\sum_n \frac{1}{\sqrt{2\omega(k_n)}}(\hat{a}_{k_n}e^{ik_n x}+\hat{a}^{\dagger}_{k_n}e^{-ik_n x}),\\
\hat{\Pi}(x)&=&-\frac{i}{\sqrt{V}}\sum_n \frac{\sqrt{\omega(k_n)}}{2}(\hat{a}_{k_n}e^{ik_n x}-\hat{a}^{\dagger}_{k_n}e^{-ik_n x}),\\
\end{array}
\]
where the dispersion relation is $\omega(k_n)=\sqrt{m_0^2+k_n^2}$. The ground state field-field and momentum/momentum correlations are obtained by Fourier transform (see e.g.  \cite{BR:04})
\begin{equation}
\begin{array}{lll}
\langle \hat{\Phi}(x)\hat{\Phi}(y)\rangle& =& \frac{1}{2V}\sum_n \frac{\cos(k_n (x-y))}{\omega(k_n)},\\
\langle \hat{\Pi}(x)\hat{\Pi}(y)\rangle& =& \frac{1}{2V}\sum_n \omega(k_n) \cos(k_n (x-y)).\\
\end{array}
\end{equation}
In the limit $V\rightarrow \infty$ the continuum correlations in the massless $(m_0=0)$ phase are:
\begin{equation}
\begin{array}{lll}
\langle \hat{\Phi}(x)\hat{\Phi}(y)\rangle& =& -\frac{\ln ((x-y)^2)}{4\pi},\\
\langle \hat{\Pi}(x)\hat{\Pi}(y)\rangle& =& -\frac{1}{2\pi (x-y)^2}.\\
\end{array}
\label{contcorrmassless}
\end{equation}
In the massive $(m_0>0)$ phase:
\begin{equation}
\begin{array}{lll}
\langle \hat{\Phi}(x)\hat{\Phi}(y)\rangle& =&\frac{1}{4\pi}\int_{-\infty}^{\infty}dk \frac{\cos(k (x-y))}{\sqrt{k^2+m_0^2}}\\
&=& \frac{1}{2\pi }K_0(m_0 |x-y|),\\
\langle \hat{\Pi}(x)\hat{\Pi}(y)\rangle& =&
\frac{1}{4\pi}\int_{-\infty}^{\infty}dk \cos(k (x-y))\sqrt{k^2+m_0^2}\\
&=& -\frac{m_0}{2\pi (x-y)}K_1(m_0 |x-y|),\\
\end{array}
\label{contcorrmassive}
\end{equation}
where $K_0$ and $K_1$ are modified Bessel functions of the second kind. There are two distinct limiting behaviours of correlations in the massive phase. For $|x-y|\gg m_0^{-1}$:
\begin{equation}
\begin{array}{lll}
\langle \hat{\Phi}(x)\hat{\Phi}(y)\rangle&\rightarrow&-\frac{e^{-m_0 |x-y|}}{\sqrt{8\pi m_0 |x-y|}},\\
\langle \hat{\Pi}(x)\hat{\Pi}(y)\rangle& \rightarrow &
\sqrt{\frac{m_0}{8\pi |x-y|^3}}e^{-m_0 |x-y|},
\end{array}
\end{equation}
whereas for $|x-y|\ll m_0^{-1}$:
\begin{equation}
\begin{array}{lll}
\langle \hat{\Phi}(x)\hat{\Phi}(y)\rangle&\rightarrow&-\frac{1}{2\pi}(\ln (\frac{m_0 |x-y|}{2})+\gamma),\\
\langle \hat{\Pi}(x)\hat{\Pi}(y)\rangle& \rightarrow &
-\frac{1}{2\pi (x-y)^2},
\end{array}
\end{equation}
where $\gamma$ is the Euler gamma constant.

\subsection{Decomposition in a wavelet basis}

The field and its conjugate can be expressed in a discrete wavelet family basis by projections onto the scaling and wavelet functions (here $r\geq 0$):
\begin{equation}
\begin{split}
\hat{\Phi}^{\scalezero}(n)&=\int dx\ \hat{\Phi}(x)s^{0}_{n}(x),~~
\hat{\Phi}^{[w]r}(n)=\int dx\ \hat{\Phi}(x)w^r_{n}(x),\\
\hat{\Pi}^{\scalezero}(n)&=\int dx\ \hat{\Pi}(x)s^{0}_{n}(x),~~
\hat{\Pi}^{[w]r}(n)=\int dx\ \hat{\Pi}(x)w^r_{n}(x).
\end{split}
\end{equation}
As a consequence of the orthonormality relations in Eq.~\ref{orthonormal}, the fields obey the following equal time commutation relations (assuming here that $0\leq r,s$):
\begin{equation}
\begin{split}
\ [\hat{\Phi}^{[s]r}(m),\hat{\Pi}^{[s]r}(m')]&= i \delta_{m,m'}\\
\ [\hat{\Phi}^{\waver{r}}(m),\hat{\Pi}^{\waver{r'}}(m')]&=i\delta_{r,r'}\delta_{m,m'}.\\
 \end{split}
 \end{equation}
The boundary Hamiltonian on a system with a UV cutoff scale $n$ with $V=L2^n$ modes is  \cite{BRSS:15, BP:13}:
\begin{equation}
\begin{array}{lll}
\hat{H}_{\rm bd}&=&\frac{1}{2}\big(\sum_{m=0}^{L2^n-1} :\hat{\Pi}^{\scale}(m)\hat{\Pi}^{\scale}(m):\\
&&+m_0^2\sum_{m=0}^{L2^n-1} :\hat{\Phi}^{\scale}(m)\hat{\Phi}^{\scale}(m):\\
&&+\sum_{m,m'=0}^{L2^n-1}:\hat{\Phi}^{\scale}(m)D^{[ss]0}_{m,m'}\hat{\Phi}^{\scale}(m'):\big),
\end{array}
\label{boundHam}
\end{equation}
where $:\hat{O}:$ indicates normal ordering of the operator $\hat{O}$ is taken.
The bulk Hamiltonian is a sum of three terms involving coupling of scale/scale DOFs, scale/wavelet DOFs, and wavelet/wavelet DOFs  \cite{BRSS:15, BP:13}:
\begin{equation}\label{bulkHam}
\hat{H}_{\rm bk}=\hat{H}_{\rm ss}+\hat{H}_{\rm ww}+\hat{H}_{\rm sw},
\end{equation}
where
\begin{equation}\label{Haminwaveletbasis}
\begin{split}
\hat{H}_{\rm ss}&=\frac{1}{2}\big(\sum_{m=0}^{L-1} :\hat{\Pi}^{\scalezero}(m)\hat{\Pi}^{\scalezero}(m):\\
&+m_0^2\sum_{m=0}^{L-1} :\hat{\Phi}^{\scalezero}(m)\hat{\Phi}^{\scalezero}(m):\\
&+\sum_{m,m'=0}^{L-1}:\hat{\Phi}^{\scalezero}(m)D^{[ss]-n}_{m,m'}\hat{\Phi}^{\scalezero}(m'):\big),\\
\hat{H}_{\rm ww}&=\frac{1}{2}\big(\sum_{m=0}^{L2^r-1}\sum_{r=0}^{n-1}:\hat{\Pi}^{[w]r}(m)\hat{\Pi}^{[w]r}(m):\\ 
&+m_0^2\sum_{m=0}^{L2^r-1}\sum_{r=0}^{n-1} :\hat{\Phi}^{[w]r}(m)\hat{\Phi}^{[w]r}(m):\\
&+\sum_{m=0}^{L2^r-1}\sum_{m'=0}^{L2^{r\prime}-1}\sum_{r,r'=0}^{n-1}:\hat{\Phi}^{[w]r}(m)D^{[ww]r-n,r'-n}_{m,m'}\hat{\Phi}^{[w]r'}(m'):\big),\\
\hat{H}_{\rm sw}&=\frac{1}{2}\sum_{m'=0}^{L-1}\sum_{m=0}^{L2^r-1}\sum_{r=0}^{n-1}:\hat{\Phi}^{[w]r}(m)D^{[sw]r-n,{-n}}_{m,m'}\hat{\Phi}^{\scalezero}(m').
\end{split}
\end{equation}
 The coupling coefficients are 
\begin{equation}
\begin{split}
D^{[ss]k}_{m,m'}&=\int dx\ \partial_x s^{k}_{m}(x)\cdot \partial_x s^{k}_{m'}(x),\\
D^{[ww]r,r'}_{m,m'}&=\int dx\ \partial_x w^r_{m}(x)\cdot \partial_x w^{r'}_{m'}(x),\\
D^{[sw]r,k}_{m,m'}&=2 \int dx\ \partial_x w^r_{m}(x)\cdot \partial_x s^{k}_{m'}(x).\\
\end{split}
\label{deriviativeoverlaps}
\end{equation} 
These can be computed systematically as described in Appendix \ref{deroverlap}.
For wavelet family $\mathcal{K}$, the smallest size admissible with periodic boundaries is $L=2(2\mathcal{K}-1)$ in order to ensure the scale and wavelet functions at the coarsest scale are orthonormal.  

As described in Appendix \ref{diagHambound}, the boundary Hamiltonian can be diagonalized as
\begin{equation}
\hat{H}_{\rm bd}=\sum_{j=0}^{V-1} d_j \left(\hat{\tilde{a}}^{\dagger}_j\hat{\tilde{a}}_j+\frac{1}{2}\right),
\end{equation}
with normal mode operators $\hat{\tilde{a}}_j=\frac{1}{\sqrt{V}}\sum_{m=0}^{V-1}e^{i 2 \pi jm/V}\hat{a}^{[s]n}(m)$ and eigenenergies
\begin{equation}
d_j=\Big[(-D^{[ss]0}_{0,0} +m_0^2)+2\sum_{m=0}^{2\mathcal{K}-2}D^{[ss]0}_{0,m}\cos(k_j m)\Big]^{1/2}.
\label{eigs}
\end{equation}
In the massless case there is a boundary zero mode for $\hat{H}_{\rm bd}$ defined by the mode operator: $\hat{\tilde{a}}^{[s]n}(0)$. The ground state is the unique state that satisfies $\hat{\tilde{a}}^{[s]n}(j)\ket{G}_{\rm bd}=0, \forall j$. 

The bulk Hamiltonian can similarly be diagonalized as
\begin{equation}
\hat{H}_{\rm bk}=\sum_{j=0}^{V-1} d_j \left(\hat{\tilde{b}}^{\dagger}_j\hat{\tilde{b}}_j+\frac{1}{2}\right),
\end{equation}
where the eigenenergies are the same as for the boundary and the normal mode operators are sums of operators that act on scale and wavelet DOFs. The bulk normal modes are
\begin{equation}
\hat{\tilde{b}}_j=\sum_{k=0}^{V-1} [M_{\rm bk}]_{j,k}\hat{b}_k,
\label{bulkmode}
\end{equation}
where the vector of bulk annihilation operators is
\begin{equation}
\begin{split}
\hat{\bm{b}}&=(\hat{a}^{[s]0}(0),\ldots, \hat{a}^{[s]0}(2L-1),\hat{a}^{[w]0}(0),\ldots, \hat{a}^{[w]0}(2L-1),\\
&~~~~~~\hat{a}^{[w]1}(0),\ldots,\hat{a}^{[w]1}(2\times 2L-1),\ldots \hat{a}^{[w]n-1}(0),\ldots,\\
&~~~~~~\hat{a}^{[w]n-1}(2^{n-1}\times 2L-1))^T,\nonumber
\end{split}
\end{equation}
and $M_{\rm bk}$ is an orthogonal wavelet transform matrix defined in Sec. \ref{Circuit}. In the massless phase, the zero mode in the bulk is a wavelet transformation of the zero mode on the boundary, which in turn is a uniform superposition of localized modes. The wavelet transform of any uniform vector has support only on the coarsest scale DOFs, and using that fact together with translational invariance of the zero mode, we have the expression for the bulk zero mode:
$\hat{\tilde{b}}_0=\frac{1}{\sqrt{2L}} \sum_{m=0}^{2L-1}\hat{a}^{[s]0}(m)$.

 \section{Bulk/boundary correspondence in the ground state}
 \label{BB}

We want to compare the properties of the ground state $\ket{G}_{\rm bk}$ of $\hat{H}_{\rm bk}$ to the ground state $\ket{G}_{\rm bd}$ of $\hat{H}_{\rm bd}$. 
The ground state $\rho$ of a quadratic bosonic Hamiltonian on $V$ modes is completely described by the covariance matrix:
\beq
\Gamma_{j,k} = \Re[ \tr[\rho (\hat{\vec{r}}_j-\langle \hat{\vec{r}}_j\rangle) (\hat{\vec{r}}_k-\langle \hat{\vec{r}}_k\rangle)]].
\eeq
Here $\langle\hat{\vec{r}}_j\rangle$ is the expectation value of $j$-th component of a $2V$ dimensional vector of field operators and their conjugate momenta. For the scalar bosonic theory, the means are zero so
\beq
\Gamma_{j,k} = \Re[ \tr[\rho \hat{\vec{r}}_j \hat{\vec{r}}_k]].
\eeq
For the boundary Hamiltonian $\hat{H}_{\rm bd}$, the ground state covariance matrix is
\[
\Gamma_{\rm bd}=\frac{1}{2} \left(\begin{array}{cc}K^{-1/2}_{\rm bd} & 0 \\0 & K_{\rm bd}^{1/2}\end{array}\right),
\]
expressed in the basis given by components of the vector
\begin{equation}
\hat{\bm{r}}_{\rm bd}=(\hat{\Phi}^{[s]n}(0),\ldots, \hat{\Phi}^{[s]n}(V-1),\hat{\Pi}^{[s]n}(0),\ldots, \hat{\Pi}^{[s]n}(V-1))^T.
\label{opvecbd}
\end{equation}
Here the boundary couplings are
\begin{equation}
[K_{\rm bd}]_{a,b}=(m_0^2-D^{[ss]0}_{0,0})\delta_{a,b}+D^{[ss]0}_{0,a\oplus_V(-b)}+D^{[ss]0}_{0,b\oplus_V(-a)}.
\end{equation}
Similarly, for the bulk Hamiltonian, $\hat{H}_{\rm bk}$, the ground state covariance matrix is \cite{BRSS:15}
\begin{equation}
\Gamma_{\rm bk}=\frac{1}{2} \left(\begin{array}{cc}K^{-1/2}_{\rm bk} & 0 \\0 & K_{\rm bk}^{1/2}\end{array}\right),
\end{equation}
expressed in the basis given by components of the vector
\begin{equation}
\begin{array}{lll}
\hat{\bm{r}}_{\rm bk} &=& (\hat{\Phi}^{[s]0}(0),\ldots, \hat{\Phi}^{[s]0}(L-1),\hat{\Phi}^{[w]0}(0)\ldots, \hat{\Phi}^{[w]0}(L-1),\\
&&\ldots, \hat{\Phi}^{[w]n-1}(0),\ldots, \hat{\Phi}^{[w]n-1}(2^{n-1}L-1),\\
&&(\hat{\Pi}^{[s]0}(0),\ldots, \hat{\Pi}^{[s]0}(L-1),\hat{\Pi}^{[w]0}(0)\ldots, \hat{\Pi}^{[w]0}(L-1),\\
&&\ldots, \hat{\Pi}^{[w]n-1}(0),\ldots, \hat{\Pi}^{[w]n-1}(2^{n-1}L-1))^T.
\end{array}
\label{opvecbk}
\end{equation}
The bulk couplings are:

\begin{equation}\label{Kmatrix}
K_{\rm bk}=\left[\begin{array}{cccc}K_{\rm ss} & K_{\rm sw}(0) & \cdots & K_{\rm sw}({n-1}) \\
\ K^T_{\rm sw}(0)  & K_{\rm ww}(0,0)  & \ldots & K_{\rm ww}(0,{n-1})  \\
\quad\vdots &  & \ddots &  \\
\ K^T_{\rm sw}({n-1})  & \dots & \cdots & K_{\rm ww}({n-1},{n-1})
\end{array}\right].
\end{equation}
The scale-scale mode couplings are encoded in $K_{\rm ss}$, the scale-wavelet couplings in $K_{\rm sw}$ and the wavelet-wavelet couplings in $K_{\rm ww}$. These matrices are:
\begin{equation}\label{BigK}
\begin{split}
&[K_{\rm ss}]_{a,b}=(m_0^2-D^{[ss]-n}_{0,0})\delta_{a,b}+D^{[ss]-n}_{0,(a-b)\bmod{L}}+D^{[ss]-n}_{0,(b-a)\bmod{L}}\\
&\quad \quad (0\leq a,b<L),\\
&[K_{\rm sw}(l)]_{a,b}=D^{[sw]-n,l-n}_{a,b}
\quad \quad (0\leq a<L, 0\leq b < L2^l, 0\leq l<n),\\
&[K_{\rm ww}(l,j)]_{a,b}=m_0^2\delta_{a,b}\delta_{j,l}+D^{[ww]l-n,j-n}_{a,b}\\
&\quad \quad (0\leq a<L2^l, 0\leq b < L2^j, 0\leq j\leq l<n).\\
\end{split}
\end{equation}

Note that any local operator on the bulk can be written as a superposition of field operators on the boundary:
\begin{equation}
\hat{A}^{\scale}(m)=\sum_j c_{n,j,m} \hat{A}^{[s]0}(j)+\sum_{j=0}^{L2^r-1} \sum_{r=0}^{n-1} d_{n,r,j,m} \hat{A}^{[w]r}(j),
\end{equation}
where for $0\leq r<n$,
\[
\begin{array}{lll}
c_{n,j,m}&=&\int dx s^0_j(x)s^n_m(x),\\
 d_{n,r,j,m}&=&\int dx w^r_j(x)s^n_m(x).
 \end{array}
 \]
Then correlation functions on the boundary can be expressed in terms of $O(n)$ correlations in the bulk:
 \begin{equation}
 \begin{array}{lll}
&_{\rm bd}\bra{G}\hat{A}^{\scale}(m)\hat{B}^{\scale}(m')\ket{G}_{\rm bd}&\\
&=\sum_{j,j'=0}^{L-1}c_{n,j,m}c_{n,j',m'} {_{\rm bk}}\bra{G}\hat{A}^{\scalezero}(m)\hat{B}^{\scalezero}(m')\ket{G}_{\rm bk}&\\
&+\sum_{j=0}^{L-1}\sum_{j'=0}^{L2^{r\prime}-1}\sum_{r'=0}^{n-1}c_{n,j,m}d_{n,r',j',m'} {_{\rm bk}}\bra{G}\hat{A}^{\scalezero}(j)\hat{B}^{[w]r'}(j')\ket{G}_{\rm bk}&\\
&+\sum_{j=0}^{L2^r-1}\sum_{j'=0}^{L-1}\sum_{r=0}^{n-1}d_{n,r,j,m}c_{n,j',m'} {_{\rm bk}}\bra{G}\hat{A}^{[w]r}(j)\hat{B}^{\scalezero}(j')\ket{G}_{\rm bk}&\\
&+\sum_{j=0}^{L2^r-1}\sum_{j'=0}^{L2^{r\prime}-1}\sum_{r,r'=0}^{n-1}d_{n,r,j,m}d_{n,r',j',m'} \\
&{_{\rm bk}}\bra{G}\hat{A}^{[w]r}(j)\hat{B}^{[w]r'}(j')\ket{G}_{\rm bk}.&\\
\label{bdcorrelationsfrombk}
\end{array}
\end{equation}

\begin{figure}[t]
	\begin{center}
		\includegraphics[width=\columnwidth]{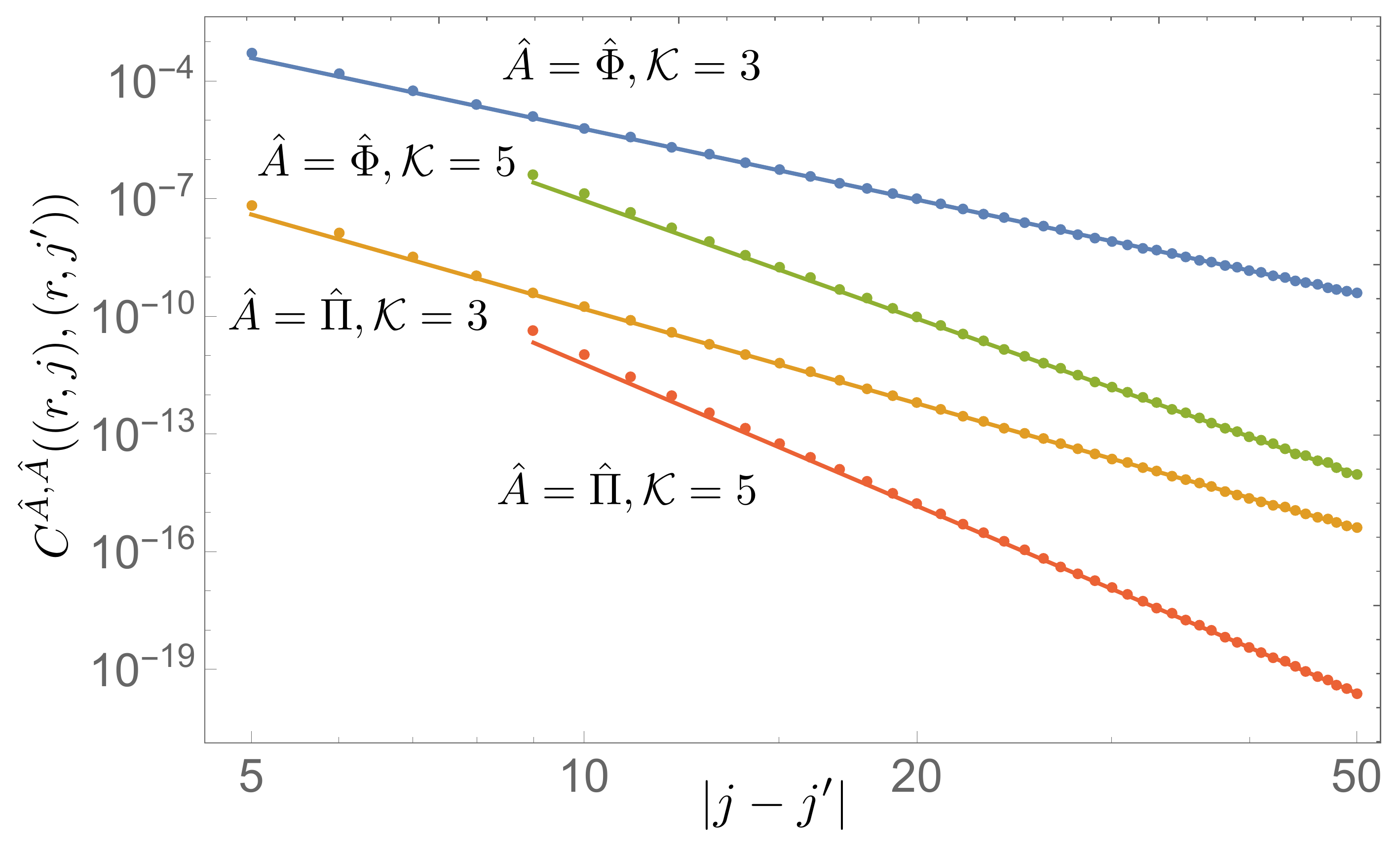}
	\end{center}
	\caption{Same scale bulk correlations as a function of position separation for a massless field theory. Here the bulk scale is $r=8$ and the boundary scale is $n=15$. Bulk field-field ($\hat{A}=\hat{\Phi})$ and momentum-momentum ($\hat{A}=\hat{\Pi}$) correlations for the $\mathcal{K}=3,5$ wavelet families are calculated using the ground state of the boundary Hamiltonian (Eq.~\ref{corrs}) (dots). Solid lines are the derived relations for the bulk correlations from Eqs.~\ref{phiphicorr},\ \ref{pipicorr}.}
		\label{fig:SameScalecorrs}
\end{figure}
We also want to compute the scaling of two point correlations in the bulk. This can be done by calculating correlations on the wavelet DOFs only:
\begin{equation}
C^{\hat{A},\hat{B}}((r,j),(r',j'))={_{\rm bk}}\bra{G}\hat{A}^{[w]r}(j)\hat{B}^{[w]r'}(j')\ket{G}_{\rm bk},
\end{equation}
where $\hat{A}$ and $\hat{B}$ are local wavelet mode operators. The bulk correlations can be calculated as follows.
For $0\leq r<n$, define
\[
f_{n,r,j,m}=\int dx s^n_m(x)w^r_j(x);\\
\]
then 
\begin{equation}
\begin{array}{lll}
C^{\hat{A},\hat{B}}((r,j),(r',j'))&=&\displaystyle{\sum_{m,m'=0}^{L2^{r}-1}} f_{n,r,j,m}f_{n,r',j',m'}\\
&& {_{\rm bd}\bra{G}}\hat{A}^{\scale}(m)\hat{B}^{\scale}(m')\ket{G}_{\rm bd}.
\end{array}
\end{equation}
For $n\gg r$, the overlap integral effectively samples the wavelet at the location of the scale field, i.e. $f_{n,r,j,m}\approx   \int dx w^r_j(\bar{x}) s^n_m(x)= 2^{-n/2}w^r_j(\bar{x})$, where $\bar{x}=2^{-n}m$ is the approximate location of the peak of the scale field $s^n_m$. Hence, for $n\gg r$, the overlap coefficients can be expressed as follows:
\[
 f_{n,r,j,m}\approx\left\{\begin{array}{cc}w_0^{r-n}(m-j2^{n-r}+1) & m\in 2^{n-r}[j,j+2\mathcal{K}-1] \bigcap\mathbb{Z}\\0 & {\rm otherwise}.\end{array}\right.
  \]
The correlator can be written for $0\leq r,r'<n$ as:
\begin{equation}
\begin{array}{lll}
C^{\hat{A},\hat{B}}((r,j),(r',j'))
&=&2^{-n}2^{(r+r')/2}\displaystyle{\sum_{m=j\times 2^{n-r}}^{(j+2\mathcal{K}-1)\times 2^{n-r}}}\sum_{m'=j'\times 2^{n-r'}}^{(j'+2\mathcal{K}-1)\times 2^{n-r'}}\\
&&w_0^{0}((m-j\times 2^{n-r}+1)\times 2^{r-n}) \\
&&w_0^{0}((m'-j'\times 2^{n-r'}+1)\times 2^{r'-n})\\
&&{_{\rm bd}\bra{G}}\hat{A}^{\scale}(m)\hat{B}^{\scale}(m')\ket{G}_{\rm bd}.\\
\end{array}
\label{corrs}
\end{equation}
Here we have used the property: $w_0^{r}(x)=2^{r/2}w_0^{0}(x\times 2^{r})$.
To find the boundary correlations we need to find the ground state of the boundary Hamiltonian $\hat{H}_{\rm bd}$ (see Appendix \ref{diagHambound}). When the number of modes is very large, the correlations in the ground state approach the continuum values: Eqs.~ \ref{contcorrmassless},\ref{contcorrmassive}.

\subsection{Massless case}
\label{massless}
\begin{figure}[t]
	\begin{center}
		\includegraphics[scale=0.4]{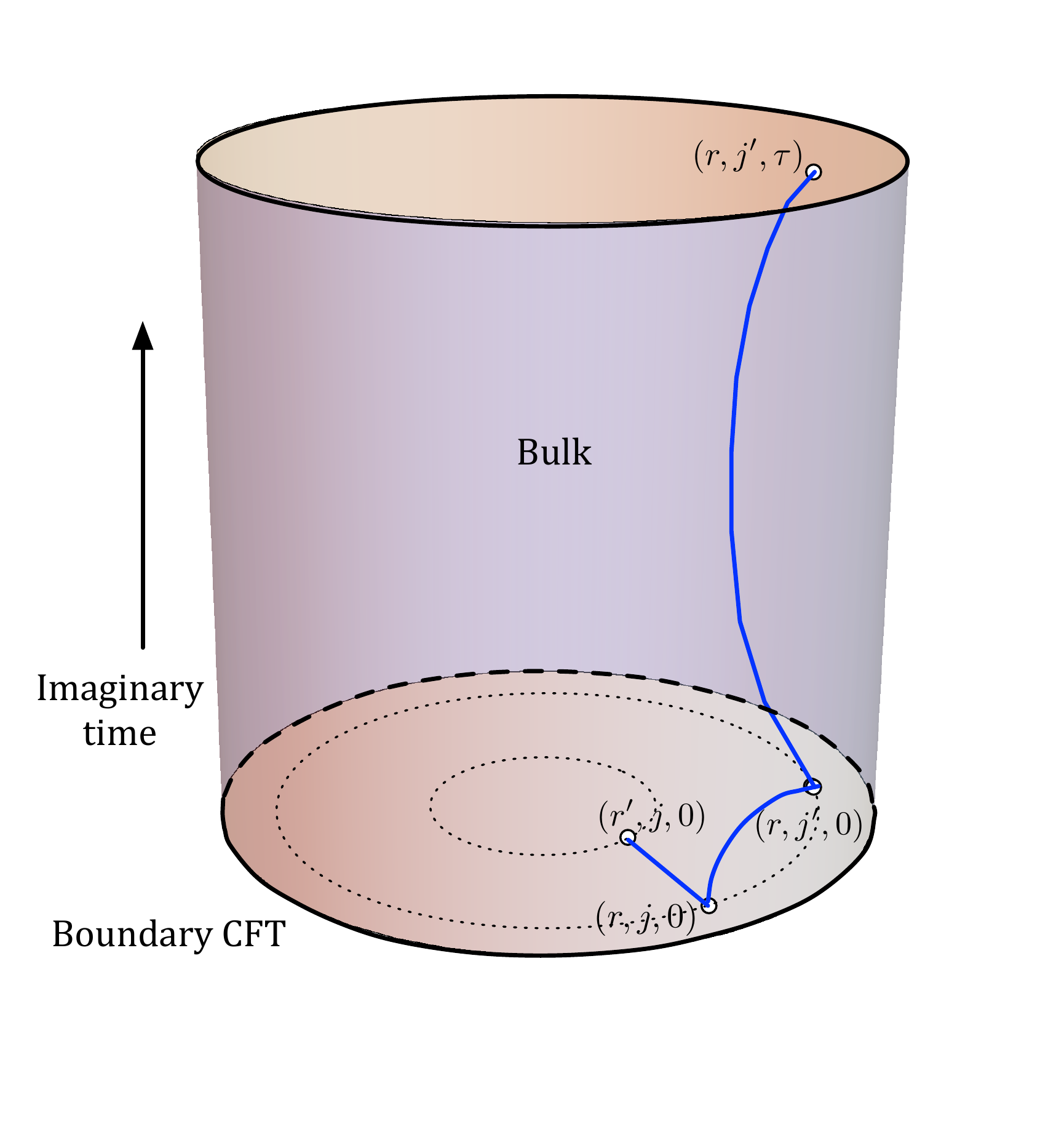}
		\label{fig:masslessopenent}
	\end{center}
	\caption{Illustration of the bulk boundary correspondence for the bosonic $1+1$ dimensional CFT flowing in imaginary time. The bulk wavelet degrees of freedom have correlations that decay exponentially with a distance given by geodesics in   AdS$_{3}$ space. The radius of curvature $R$ of the emergent bulk geometry depends on the wavelet family $\mathcal{K}$. Blue lines depict geodesic distances between space time points.}
\end{figure}
The bulk correlations can be computed from the boundary correlations. As derived in Appendix \ref{Bkcorrm0}, the same scale bulk correlations for separation $j>2\mathcal{K}-1$ and deep in the bulk are
\begin{equation}
C^{\hat{\Phi},\hat{\Phi}}((r,0),(r,j))\approx -\frac{2^{n-r} \times D_{\mathcal{K}}}{4\pi \mathcal{K}j^{2\mathcal{K}}},
\label{phiphicorr}
\end{equation}
where
\begin{equation}
D_{\mathcal{K}}=\langle x^{\mathcal{K}}\rangle_w^2{2\mathcal{K} \choose \mathcal{K}},
\end{equation}
and $\langle x^{\mathcal{K}}\rangle_w$ is the $\mathcal{K}$th moment of the wavelet $w_0^0(x)$.
Similarly, the momentum-momentum correlations are
\begin{equation}
C^{\hat{\Pi},\hat{\Pi}}((r,0),(r,j))\approx \frac{2^{r-n}\times (2\mathcal{K}+1)D_{\mathcal{K}}}{2\pi j^{2\mathcal{K}+2}},
\label{pipicorr}
\end{equation}
For example, using the wavelet moments calculated in Appendix \ref{momentscalc}: $D_3=\frac{225}{128}, D_4=21.53, D_5=446.04$. This fits the numerically calculated correlations in Fig. \ref{fig:SameScalecorrs} quite well.
Additionally we find for self correlations
\begin{equation}
C^{\hat{\Phi},\hat{\Phi}}((r,0),(r,0))= 2^{n-r-a},
\end{equation}
and
\begin{equation}
C^{\hat{\Pi},\hat{\Pi}}((r,0),(r,0))= 2^{r-n+b}.
\end{equation}
The values $a,b$ appearing in the self correlations can be calculated from the boundary correlations (see Appendix \ref{samescalecorrs}).
For $\mathcal{K}\geq 3$, $a\approx 3.18$ independent of $\mathcal{K}$, whereas $b\approx 0.75/\mathcal{K}^2+0.16/ \mathcal{K}+1.24$.

\subsubsection{Mutual Information}
The mutual information between two bulk DOFs is given by 
\begin{equation}
I((r,j),(r',j'))=S(\rho_{(r,j)})+S(\rho_{(r',j')})-S(\rho_{(r,j),(r',j')}),
\end{equation}
where the von Neumann entropy $S(\rho_A)$ of a subsystem $A$ is calculated as follows. 
First obtain the reduced covariance matrix $\Gamma_A$ by deleting the columns and rows of modes not contained in $A$ from the full covariance matrix $\Gamma$. Next compute the symplectic spectrum, which are the eigenvalues of the matrix $i\Gamma_A\Omega$ where $\Omega$ is the symplectic form
\begin{equation}
\Omega=\left(\begin{array}{cc}0 & {\bf 1}_{n_A} \\-{\bf 1}_{n_A} & 0\end{array}\right),
\label{BigOmega}
\end{equation}
written the basis $\{\hat{\Phi}_1,\ldots, \hat{\Phi}_{n_A},\hat{\Pi}_1,\ldots, \hat{\Pi}_{n_A}\}$ of the $n_A$ modes of system $A$.
These eigenvalues come in positive and negative pairs $\{\pm \sigma_i\}$. Taking the positive values, the entropy in bits is 
\[
S(\rho_A)=\sum_{\{\sigma_i\}}[(\sigma_i+\frac{1}{2})\log_2 (\sigma_i+\frac{1}{2})-(\sigma_i-\frac{1}{2})\log_2 (\sigma_i-\frac{1}{2})].
\]
\begin{figure}[t]
	\begin{center}
		\includegraphics[width=\columnwidth]{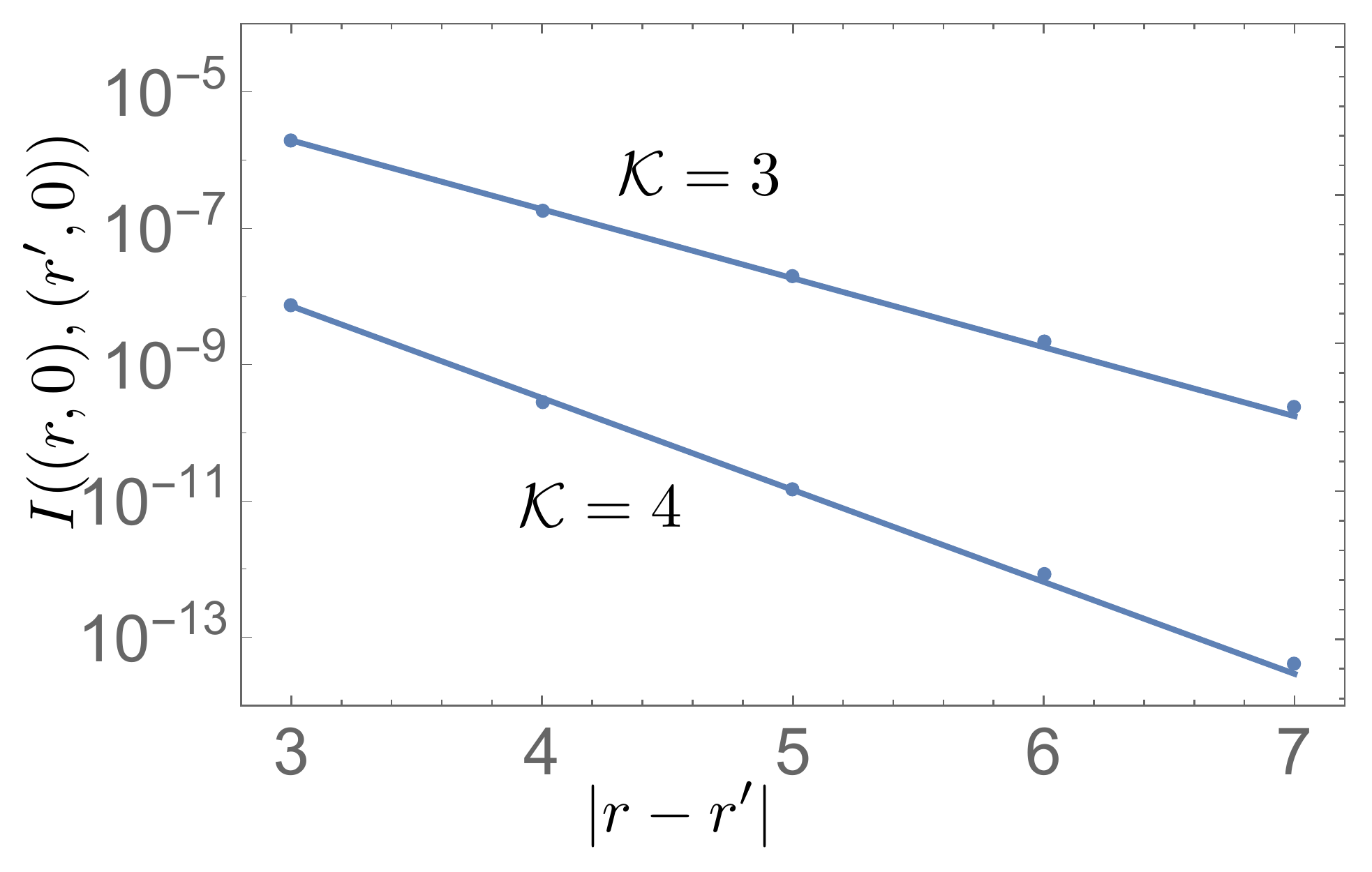}
	\end{center}
	\caption{Cross scale bulk mutual information as a function of scale separation for a massless field theory. Here $r'<r=14$ and the boundary scale is $n=15$. Mutual information is computed from boundary correlations  using Eq.~\ref{corrs} (dots), and the function $I_0e^{-d_g((r,0),(r',0))/\xi_r}$ is plotted (solid line) with the fit $\xi_r=1.79 \xi_{\theta}$.
}
	\label{fig:CrossScalecorrs}
\end{figure}

From the derived values for the same scale correlations, we find the positive symplectic eigenvalues for the subsystem pair of sites $A=\{(r,0),(r,j)\}$
\[
\sigma_{\pm}=\frac{2^{-a/2}\sqrt{j^{2\mathcal{K}}\pm 2^a \frac{D_{\mathcal{K}}}{4\pi \mathcal{K} }}\sqrt{2^b j^{2\mathcal{K}+2}\mp \frac{(2\mathcal{K}+1)D_{\mathcal{K}}}{2\pi}}}{j^{2\mathcal{K}+1}}.
\]
while for the single mode subsystem $A=\{(r,0)\}$, the positive symplectic eigenvalue is
\[
\sigma=2^{(b-a)/2}.
\]
To leading order in $1/j$, for $j>2\mathcal{K}-1$, the mutual information between bulk sites at the same scale $r$ is
\begin{equation}
I((r,0),(r,j))=\frac{\left(\frac{D_{\mathcal{K}}}{4\pi \mathcal{K}}\right)^2 F(\mathcal{K})}{j^{4\mathcal{K}}},
\label{mutualinfo}
\end{equation}
where
\[
\begin{array}{lll}
F(\mathcal{K})&=&2^{2a-2}S_0+\frac{(2^{2-2a}-2^{-a-b})^{-1}}{\log (2)}-\frac{2^{2a-3}\log(2^{b-a}-\frac{1}{4})}{\log 2}.\\
\end{array}
\]
the $\mathcal{K}$ dependence arising since $b$ is a function of $\mathcal{K}$.
Here the single site entropy $S_0$ is
\begin{equation}
\begin{array}{lll}
S_0\equiv S(\rho_{(r,j)})&=&(2^{(b-a)/2}+\frac{1}{2})\log_2(2^{(b-a)/2}+\frac{1}{2})\\
&&-(2^{(b-a)/2}-\frac{1}{2})\log_2(2^{(b-a)/2}-\frac{1}{2}).
\end{array}
\end{equation}
By translational invariance, deep in the bulk, $S_0$ is the same at any position.

\subsubsection{Computing the radius of curvature}
\label{RadiusComp}
Following Ref. \cite{Qi:13}, we make the ansatz that the mutual information falls off exponentially with the distance between two bulk modes in the AdS$_3$ metric, i.e. $I((r,0),(r,j))=S_0 e^{-d_g((r,0),(r,j)/\xi_{\theta}}$ with geodesic distance $ d_g((r,0),(r,j))= 2R\ln\left[\frac{j}{R}\right]$, and $\xi_{\theta}$ a correlation length. The multiplicative constant $S_0$ is used since for zero separation, the mutual information is just the single site entropy.  Taking logarithms of the mutual information,
\[
\ln S_0 -\frac{2R\ln j}{\xi_{\theta}}+\frac{2R\ln R}{\xi_{\theta}}=\ln \left[\left(\frac{D_{\mathcal{K}}}{4\pi \mathcal{K}}\right)^2F(\mathcal{K})\right]-4\mathcal{K}\ln j.
\]
Hence the correlation length is 
\begin{equation}
\xi_{\theta}=\frac{R}{2\mathcal{K}},
\label{samescalecorrlength}
\end{equation}
and the the radius of curvature is
\begin{equation}
R=\left( \frac{D_{\mathcal{K}}}{4\pi \mathcal{K}}\right)^{1/2\mathcal{K}}\times \left(\frac{F(\mathcal{K})}{S_0} \right)^{1/4\mathcal{K}}.
\end{equation}
Calculating the wavelet moments, for $\mathcal{K}\geq 3$ we find
\[
\left(\frac{D_{\mathcal{K}}}{4\pi \mathcal{K}}\right)^{1/2\mathcal{K}}\approx 0.319\mathcal{K}-0.357.
\]
\begin{figure}[t]
	\begin{center}
		\includegraphics[scale=0.4]{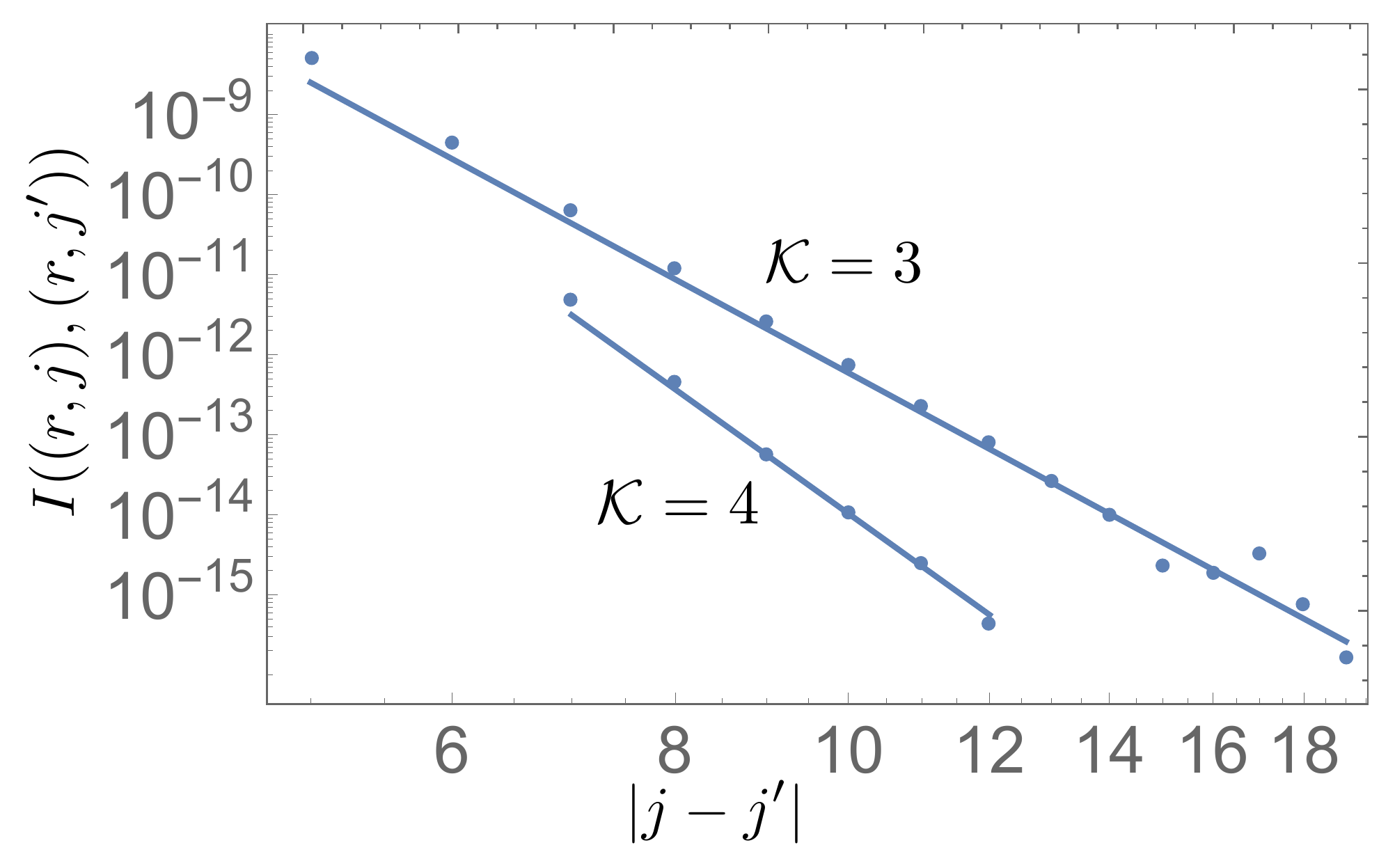}
	\end{center}
	\caption{Mutual information between same scale bulk DOFs as a function of position separation for a massless field theory. Here the bulk scale is $r=8$ and the boundary scale is $n=15$. The dots are computed values from the boundary correlations for the $\mathcal{K}=3,4$ wavelet families and solid lines are the prediction assuming exponential decay in geodesic distance $I=S_0 e^{-d_g((r,0),(r,j))/\xi_{\theta}}$ with the radius of curvature $R$ from Eq.~\ref{radius} and correlation length $\xi_{\theta}$ from Eq.~\ref{samescalecorrlength}. For $\mathcal{K}=3$: $R=1.10$, $\xi_{\theta}=0.18$, $S_0=0.22$. For $\mathcal{K}=4$: $R=1.49$, $\xi_{\theta}=0.19$, $S_0=0.18$. }
	\label{fig:mutualinfo}
\end{figure}
As described in Sec. \ref{massless}, the values appearing in the self correlators are $a= 3.18$ and $b= 0.75/\mathcal{K}^2+0.16/\mathcal{K}+1.24$, yielding
\[
\left(\frac{F(\mathcal{K})}{S_0}\right)^{1/4\mathcal{K}}\approx \frac{2.41}{\mathcal{K}}+1 .
\]
Hence for $\mathcal{K}\geq 3$ the radius of curvature is approximately
\begin{equation}
R\approx (0.32\mathcal{K}-0.88/\mathcal{K}+0.43).
\label{radius}
\end{equation}
In Figs.~\ref{fig:CrossScalecorrs},\ref{fig:mutualinfo} we plot mutual information for the wavelet families $\mathcal{K}=3,4$ which shows a good match with the above calculated value for the radius of curvature.

The above scaling of $R$ indicates that the geometry becomes flatter with larger $\mathcal{K}$ while the same scale correlation length approaches a constant $\xi_{\theta}\rightarrow 0.16$.
This is consistent with the feature that correlations fall off faster in the bulk when using wavelet transformations with larger $\mathcal{K}$, since for each stage of renormalization of the boundary state, more short range entanglement is removed. The wavelet packet transform for $\mathcal{K}$ wavelets can be implemented using a circuit of nearest neighbour mode couplings repeated $\mathcal{K}$ times \cite{EW:16b}. This can explains why the radius of curvature is linear in $\mathcal{K}$ since one may view each renormalization step for large $\mathcal{K}$ wavelets as a linear in $\mathcal{K}$ sequence of number of nearest neighbour circuits implementing $\mathcal{K}=2$ (e.g. Haar) wavelet transformations.

\subsubsection{Temporal Correlations}
\begin{figure}[t]
	\begin{center}
		\includegraphics[width=\columnwidth]{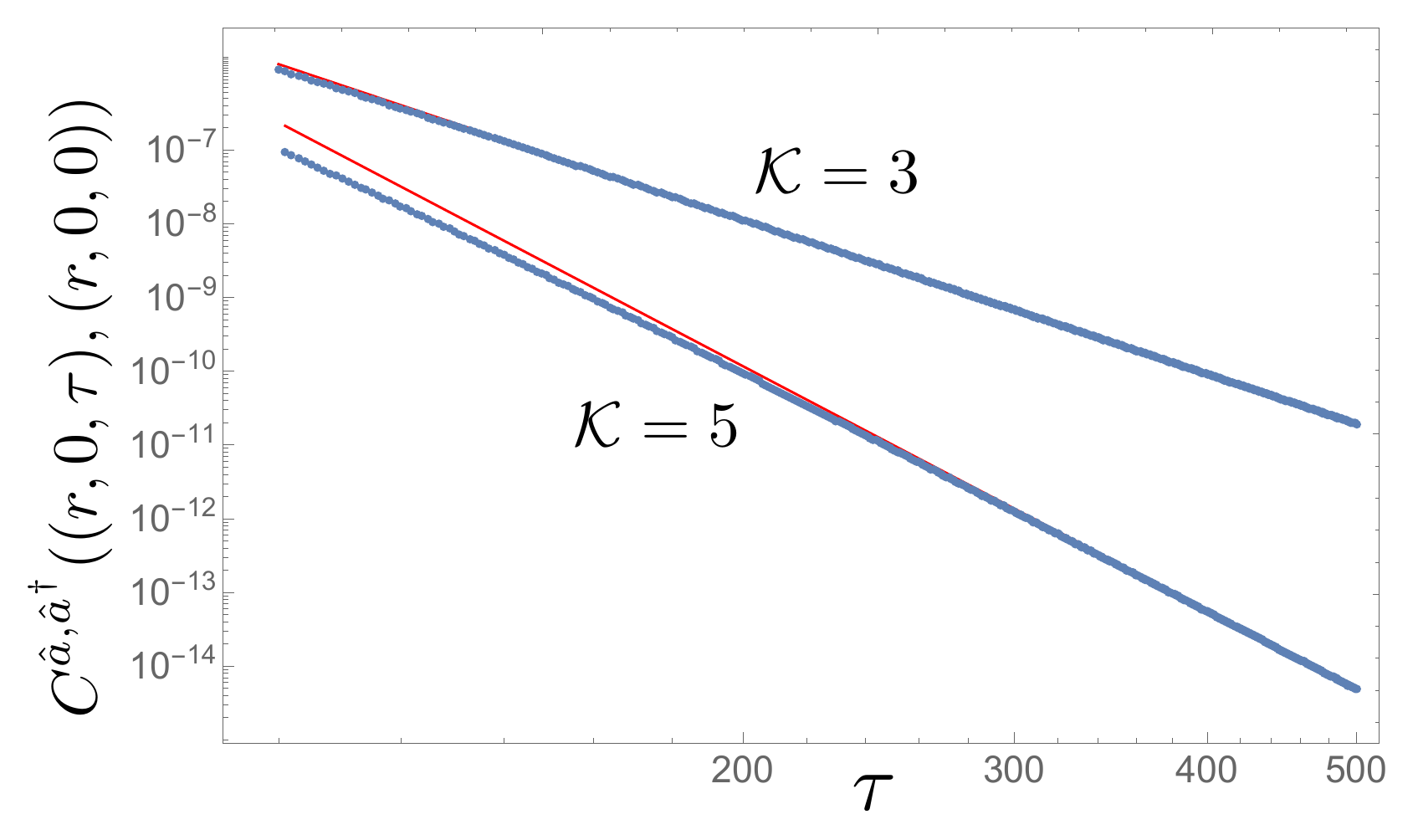}
		\label{fig:masslessopenent}
	\end{center}
	\caption{Correlation between a point in the bulk and that same point at a later imaginary time $\tau$. Blue dots are computed from the boundary theory, Eq.~\ref{fulltimecorrs} and the red lines are the prediction from Eq.~\ref{predicted}. In both cases the boundary scale is $n=7$ and the bulk DOF is at scale $r=3$ with size $L=10$. Based on Eqs.~\ref{radius}, \ref{tempcorrlen}, for $\mathcal{K}=3$ the predicted values use a radius of curvature of $R=1.10$ and $\xi_{\tau}=0.31$, whereas for $\mathcal{K}=5$, $R=1.85$ and $\xi_{\tau}=0.34$. Plots are made over the points where the predicted form is expected to be valid: $\tau>2^{n-r}(2\mathcal{K}-1)$.}
\end{figure}

To probe correlations in the time direction of the emergent bulk geometry we consider correlators of the form
\begin{equation}
\begin{split}
&C^{\hat{A},\hat{B}}((r,j,\tau),(r',j',\tau'))= {_{\rm bk}\bra{G}}T[\hat{A}^{[w]r}(m,\tau)\hat{B}^{[w]r'}(m',\tau')]\ket{G}_{\rm bd}\\
&=\displaystyle{\sum_{m,m'=0}^{L2^{r}-1}} f_{n,r,j,m}f_{n,r',j',m'}
{_{\rm bd}\bra{G}}T[\hat{A}^{\scale}(m,\tau)\hat{B}^{\scale}(m',\tau')]\ket{G}_{\rm bd},
\end{split}
\label{fulltimecorrs}
\end{equation}
where $T$ is the time ordering operator. For simplicity we work in imaginary time, $\tau\rightarrow i\tau$, and consider the Green's function at the same point in space. For $\tau>2^{n-r}(2\mathcal{K}-1)$, we find (see Appendix \ref{apptempcorrs}), 
\begin{equation}
C^{\hat{a},\hat{a}^{\dagger}}((r,j,\tau),(r,j,0))\approx \frac{2^{(n-r)(2\mathcal{K}+1)}0.32\times  D_{\mathcal{K}}}{\tau^{(2\mathcal{K}+1)}}.
\label{predicted}
\end{equation}
As before, we make the ansatz $C^{\hat{a},\hat{a}^{\dagger}}((r,j,\tau),(r,j,0))=C_0e^{-d_g((r,j,\tau),(r,j,0))/\xi_{\tau}}$ where $\xi_{\tau}$ is a temporal correlation length and the geodesic distance is approximately $d_g((r,j,\tau),(r,0,0))=2R\ln\left[\frac{\tau 2^{r-n}}{R}\right]$ (see Appendix \ref{AdSDistance}) . This approximation is valid when $2^n L \gg \tau\gg 2^{n-r} R$. Taking logarithms we find
\[
\begin{array}{lll}
\ln(C_0)-\frac{2R}{\xi_{\tau}}(\ln \tau -(n-r)\ln 2 -\ln R)&=&(n-r)(2\mathcal{K}+1)\ln 2 \\
&+&\ln G_{\mathcal{K}}-(2\mathcal{K}+1)\ln \tau.
\end{array}
\]
From the $\tau$ dependent term we find the correlation length
\begin{equation}
\xi_{\tau}=\frac{2R}{2\mathcal{K}+1}.
\label{tempcorrlen}
\end{equation}
Note this is larger than same scale spatial correlation length Eq.~\ref{samescalecorrlength}.
 
\subsubsection{Central Charge}
The bulk/boundary correspondence also allows for computing properties which depend on long range entanglement properties, such as the central charge $c$, which for the bosonic CFT is $c=1$.
The boundary central charge can be obtained from the purity of a subsystem $A$ on the boundary consisting of an interval of $\ell$ modes \cite{CC:09}:
\begin{equation}
\tr[\rho^2_A]=C \left[\frac{V}{\epsilon}\sin\left(\frac{\pi\ell}{V}\right)\right]^{-c/4},
\end{equation}
where $V$ is the total number of modes on the boundary (with periodic boundary conditions), $C$ is a constant and $\epsilon$ is an ultraviolet cutoff size. For Gaussian states, the purity of a subsystem consisting of $\ell$ modes is
\[
\tr[\rho^2_A]=\frac{1}{2^{\ell} \sqrt{{\rm det}\Gamma_A}},
\]
where $\Gamma_A$ is the covariance matrix for modes in region $A$. This implies that for $\ell\ll V$, 
\[
\ln[2^{2\ell}{\rm det}\Gamma_A]=\frac{c}{2}\ln(\ell)+{\rm const.}.
\]
If we consider two regions, $A_1$ and $A_2$ of lengths $\ell_1$ and $\ell_2>\ell_1$, then 
\begin{equation}
c=\frac{2}{\ln(\ell_2/\ell_1)}\left(\ln\left[\frac{{\rm det}\Gamma_{A_2}}{{\rm det}\Gamma_{A_1}}\right]+2(\ell_2-\ell_1)\ln 2\right).
\label{simplecc}
\end{equation}
 Since the elements of the covariance matrix consist of boundary correlations this data can be obtained from bulk correlations using Eq.~\ref{bdcorrelationsfrombk}.
Computing the central charge as per Eq.~\ref{simplecc} for the massless boundary CFT at scale $n=7$ and $L=10$, with a total number of modes $V=L\times 2^n=1280$, we find for region sizes $\ell_2=2\ell_1=6$, that $c=0.997$. 

\subsection{Massive Case}


For $m_02^{n-r}\gg 1$, the same scale field/field correlation deep in the bulk and for separations $j\gg 2\mathcal{K}-1$ is (see Appendix \ref{massive}) 
\begin{equation}
C^{\hat{\Phi},\hat{\Phi}}((r,0),(r,j))
\approx -\frac{2^{n-r}e^{-j \tilde{m}}}{\sqrt{j 8\pi \tilde{m}}}\langle e^{-\tilde{m} x}\rangle_w \langle e^{\tilde{m}x}\rangle_w.\\
\label{PhiPhimassive}
\end{equation} 
Similarly, the momentum/momentum field correlator is
\begin{equation}
C^{\hat{\Pi},\hat{\Pi}}((r,0),(r,j))
\approx 2^{r-n}e^{-j \tilde{m}}\sqrt{\frac{\tilde{m}}{8\pi j^3}}\langle e^{-\tilde{m} x}\rangle_w \langle e^{\tilde{m} x}\rangle_w.
\label{PiPimassive}
\end{equation} 
For $\mathcal{K}=3$ wavelets and $\tilde{m}\gg 1$ the product of averages is $\langle e^{-\tilde{m} x}\rangle_w \langle e^{\tilde{m} x}\rangle_w\approx e^{4.71 \times \tilde{m}-20.38}$.
We see that the same scale bulk correlations fall off exponentially with a renormalized mass
\[
\tilde{m}=m_0 2^{n-r}.
\]
This is consistent with the view that the boundary DOFs are renormalized over $n-r$ dyadic steps to the wavelet DOFs at the scale $r$.  

\section{Circuit constructions of the bulk and boundary states}
\label{Circuit}
\subsection{Ground states}
The boundary ground state is obtained by a unitary transformation on the $V$ mode vacuum state:
\[
\ket{G}_{\rm bd}=\hat{U}_{\rm bd}\ket{{\rm vac}}^{\otimes V}.
\]
 This mapping is described by a symplectic transformation on the initially decoupled position and momentum mode operators:
$\hat{\bm{r}}_{\rm bd}\rightarrow Y_{\rm bd}\hat{\bm{r}}_{\rm bd}$. 
The initial vacuum correlation functions are described by a covariance matrix proportional to the identity and the symplectric transformation acts on the correlation matrix as
\begin{equation}
\Gamma_{\rm vac}=\frac{1}{2}\bm{1}_{2V}\rightarrow \Gamma_{\rm bd}=\frac{1}{2}Y_{\rm bd}Y_{\rm bd}^T=\frac{1}{2} (K_{\rm bd}^{-1/2}\oplus K_{\rm bd}^{1/2}).
\end{equation}
Similarly for the bulk ground state, the transformation is
\[
\ket{G}_{\rm bk}=\hat{U}_{\rm bk}\ket{{\rm vac}}^{\otimes V},
\]
which corresponds to the symplectic transformation $\hat{\bm{r}}_{\rm bk}\rightarrow Y_{\rm bk}\hat{\bm{r}}_{\rm bk}$,
acting on the covariance matrix as
\begin{equation}
\Gamma_{\rm vac}=\frac{1}{2}\bm{1}_{2V}\rightarrow \Gamma_{\rm bk}=\frac{1}{2}Y_{\rm bk}Y_{\rm bk}^T=\frac{1}{2} (K_{\rm bk}^{-1/2}\oplus K_{\rm bk}^{1/2}).
\end{equation}
There is a canonical Bloch-Messiah decomposition for the unitaries $\hat{U}_{\rm bd}$ and $\hat{U}_{\rm bk}$ that can be written as one round of beam splitters and phase shifters, followed by parallel single mode squeezing, followed by a second round of beam splitters and phase shifters~\cite{Braunstein}. The decomposition is efficient, costing $O(V^2)$ elementary operations.
For the construction of the boundary state:
\[
\hat{U}_{\rm bd}=\hat{R}_{\rm bd} \hat{D}\hat{R}^{\dagger}_{\rm bd},
\]
where the the single mode squeezing operations are
\[
\hat{D}=\prod_{j=0}^{V-1} e^{\alpha_j (\hat{a}_j^{2}-\hat{a}_j^{\dagger 2})/2},
\]
with squeezing parameters
\begin{equation}
\alpha_j=-\frac{1}{4}\log(d_j).
\label{squeezing}
\end{equation}
Here the $d_j$ are eigenenergies given in Eq.~\ref{eigs}, and positive values of $\alpha$ perform momentum squeezing: $\hat{D} \hat{q}_j \hat{D}^{\dagger}=e^{\alpha_j}\hat{q}_j$, $\hat{D} \hat{p}_j \hat{D}^{\dagger}=e^{-\alpha_j}\hat{p}_j$
. The unitary $\hat{R}_{\rm bd}$ performs the linear transformation: $\hat{R}_{\rm bd}\hat{a}_j \hat{R}^{\dagger}_{\rm bd}=\sum_{k=0}^{V-1} [M_{\rm bd}]_{j,k} \hat{a}_k$, where where $M_{\rm bd}$ is an orthogonal matrix that diagonalizes $K_{\rm bd}$. Specifically, we can take (assuming $V$ even)
\[
[M_{\rm bd}]_{j,k}=\left\{\begin{array}{c}\frac{1}{\sqrt{V}}\quad j=0 \\ \sqrt{\frac{2}{V}}\cos(\frac{2\pi j k}{V})\quad 1\leq j\leq V/2-1 \\ \frac{(-1)^k}{\sqrt{V}}\quad j=V/2 \\ \sqrt{\frac{2}{V}}\sin(\frac{2\pi j k}{V})\quad V/2< j\leq V-1  \end{array}\right..
\]

Similarly, for the bulk state construction:
\[
\hat{U}_{\rm bk}=\hat{R}_{\rm bk} \hat{D} \hat{R}^{\dagger}_{\rm bk}.
\]
The unitary $\hat{R}_{\rm bk}$ performs the linear transformation: $\hat{R}_{\rm bk}\hat{a}_j \hat{R}^{\dagger}_{\rm bk}=\sum_{k=0}^{V-1}[M_{\rm bk}]_{j,k} \hat{a}_k$, where $M_{\rm bk}$ is the orthogonal wavelet transformation that diagonalizes $K_{\rm bk}$.  These transformed operators define the bulk normal modes $\hat{\tilde{b}}_j$ introduced in Eq.~\ref{bulkmode}. In the massless case, there is a free mode with energy $d_0=0$ which, since it is completely delocalized, requires an infinite amount of squeezing. However, in the wavelet basis this mode is mapped to the IR fixed point at the coarsest scale DOFs and is decoupled from the wavelet DOFs. Hence one can ignore this mode if only interested in bulk wavelet correlations. The maximum amount of squeezing is then dictated by the lowest, positive definite energies $d_1=d_{V-1}$. Owing to the relation $\sum_{m=0}^{2\mathcal{K}-2} m^2 D^{[ss]0}_{0,m}=-1$ (see Appendix \ref{deroverlap}), in the massless case, $d_1=\frac{2\pi}{V}$ and the maximum squeezing needed for a $V$ mode system is $\alpha_{\rm max}=\frac{1}{4}\log(V/2\pi)$.  
\begin{figure}[t]
	\begin{center}
		\includegraphics[width=\columnwidth]{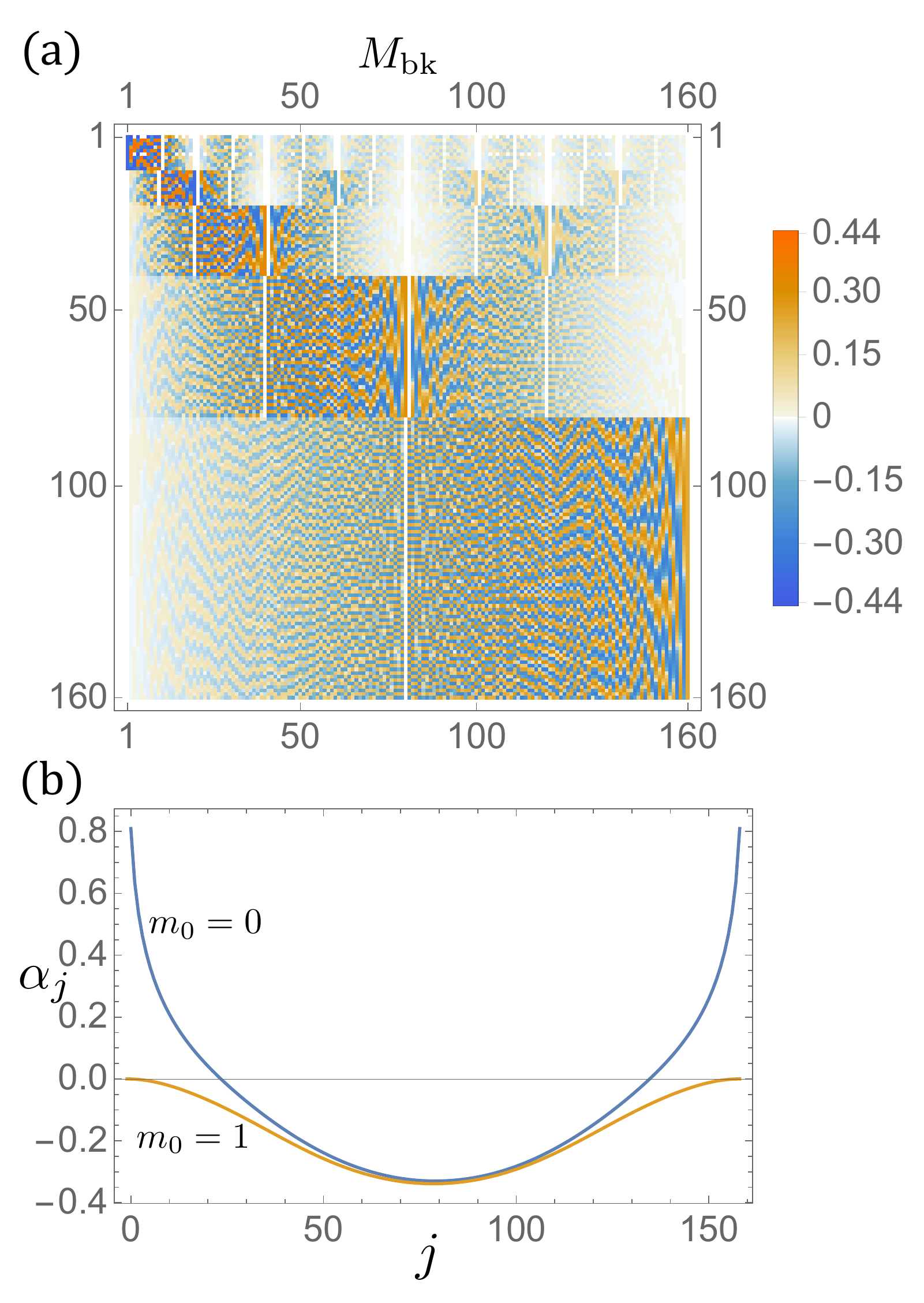}
	\end{center}
	\caption{(a) Orthogonal wavelet transformation matrix $M_{\rm bk}$ used to construct the bulk state of a system consisting of $V=160$ modes with $L=10$ and boundary scale of $n=3$. The bulk DOFs at scales $r=0,1,2,3$ are evident from the recursive structure of the matrix, with the coarsest scale DOFs in the upper left hand block. If matrix elements of magnitude less than $10^{-4}$ are set equal to zero then $M_{\rm bk}$ is $87\%$ sparse. (b) Squeezing parameters $\alpha_j$ from Eq.~\ref{squeezing} for constructing ground states in the massless case (blue) and massive $m_0=1$ case (yellow). For this sized system, the maximum momentum squeezing on a mode in the massless case is $7.03$ dB (using $\# {\rm dB}=20 \alpha \log_{10}(e)$).}
	\label{fig:FigswithTransformMatrices}
\end{figure}
\subsection{Thermal states}
The same protocol can be used to prepare thermal bulk and boundary states. 
A thermal state of the scalar bosonic QFT at temperature $1/\beta$ is described by a Gaussian state both on the bulk and on the boundary. The covariance matrices assume a simple form (see e.g. \cite{FKM:65}) on the boundary
\[
\Gamma_{\rm bd}(\beta)=\frac{1}{2} \left(\begin{array}{cc}K^{-1/2}_{\rm bd}\coth(\beta K^{1/2}_{\rm bd}) & 0 \\0 & K^{1/2}_{\rm bd}\coth(\beta K^{1/2}_{\rm bd})\end{array}\right),
\]
and on the bulk
\[
\Gamma_{\rm bk}(\beta)=\frac{1}{2} \left(\begin{array}{cc}K^{-1/2}_{\rm bk}\coth(\beta K^{1/2}_{\rm bk}) & 0 \\0 & K^{1/2}_{\rm bk}\coth(\beta K^{1/2}_{\rm bk})\end{array}\right).
\]
In the limit $\beta\rightarrow \infty$ these approach the ground state covariance matrices.
The method to construct thermal states is very similar to that for ground states. The only difference is that rather than beginning in the vacuum state one should start in the separable thermal state with correlation matrix 
\[
\Gamma(\beta)=\frac{1}{2}[\oplus_{j=0}^{V-1} \coth(\beta d_j)]\oplus [\oplus_{j=0}^{V-1} \coth(\beta d_j)].
\]
Such a state can be prepared by initializing the system in a product state such that each mode $j$ is prepared in a locally thermal state with temperature $T_j=(\beta d_j)^{-1}$. 

\section{Implemetations}
\label{Implementations}
The multimode Gaussian entangled states that are needed to demonstrate the bulk/boundary correspondence considered here can be prepared in a variety of engineered quantum systems. 
In continuous variable optical networks, large-scale Gaussian states encoded in either frequency modes~\cite{Wang:2014im,Chen:2014jx} or temporal modes~\cite{Yokoyama:2013jp} have been demonstrated, with the latter experiment realizing a $10000$ mode Gaussian cluster state. Squeezing levels of $\sim 5$dB are achievable with current technology~\cite{Yokoyama:2013jp}, and proof-of-principle experiments show squeezing of up to $\sim$10~dB ~\cite{Mehmet:2011}\cite{Yokoyama:2013jp}. 

Another available technology is circuit-QED setups using coupled arrays of microwave cavities~\cite{PazSilva:2009fa}. Single-mode~\cite{Yurke:1988ih,CastellanosBeltran:2008cg} squeezing has already been demonstrated in these systems, and the SQUID-based controlling technology allows for very strong nonlinearities enabling high squeezing ($\sim$13~dB)~\cite{Li:2011he}.

A third candidate technology is cold trapped ions. In Ref. \cite{Lau:2012} it was shown how to perform universal bosonic simulators using an array of trapped ions where each ion is trapped in its own local potential. Single model squeezing and phase shifting operations can be done by dynamically changing local trapping potentials and beam splitter operators are enabled by making use of the Coulomb repulsion between neighboring ions brought together using time dependent pairwise potentials. Additionally, thermal motional state engineering of a single ion trapped in a harmonic well has been achieved by changing the detuning of the cooling laser \cite{Itano:97} which would allow for preparation of thermal state QFT simulators.

Large sized simulations can run into difficulties addressing individual modes with local gate operations. Fortunately, as shown in Ref.~\cite{PBT:09}, the entire simulation described here consisting of state preparation, evolution by the linear optical unitaries, and measurement of quadrature moments on each mode can be done without needing mode addressability and using translationally invariant single and nearest neighbour pairwise interactions between modes arranged on a line with open boundaries. The overhead for using translationally invariant operations is only linear in the number of modes.

\begin{figure}[t]
	\begin{center}
		\includegraphics[width=\columnwidth]{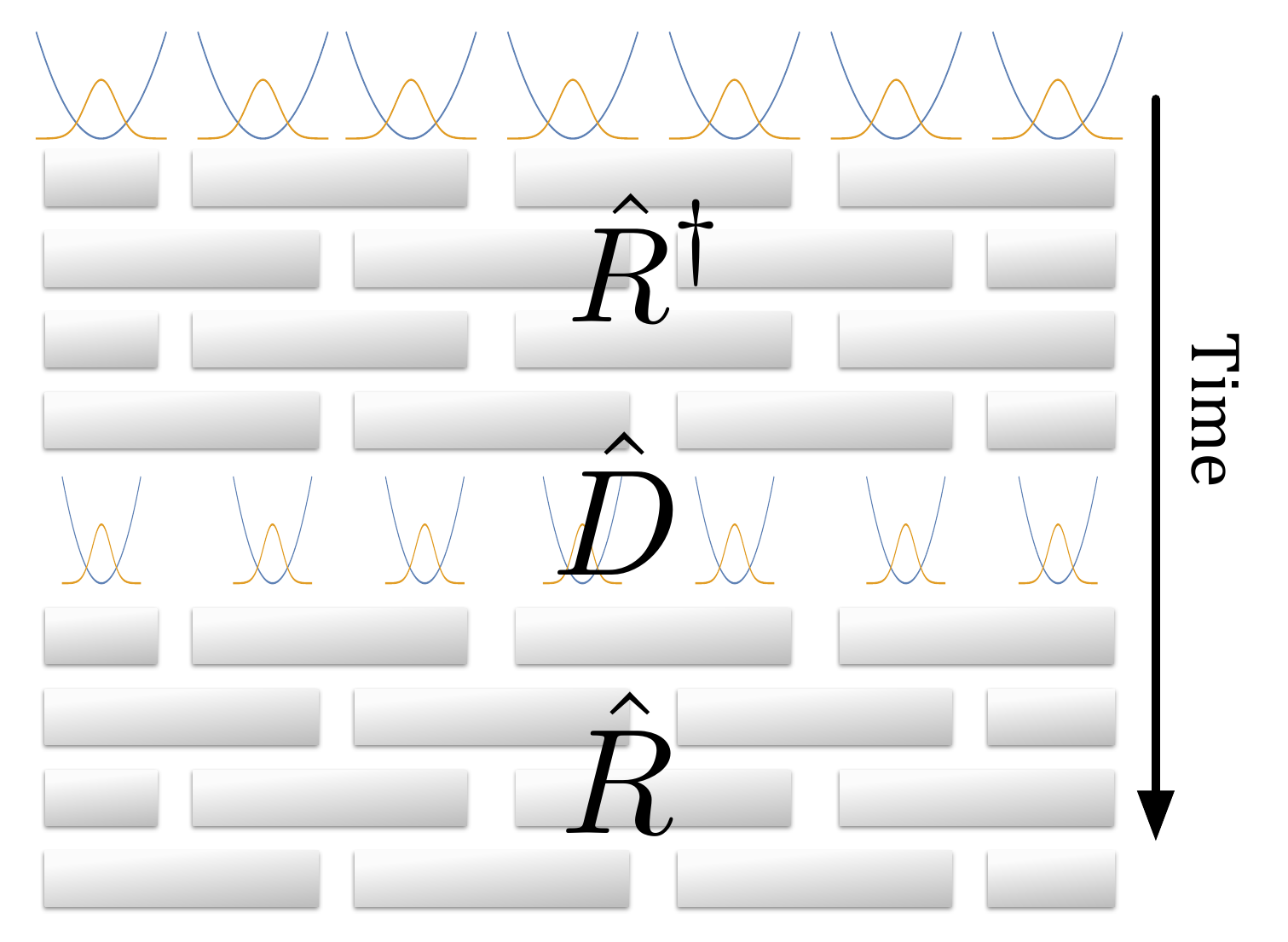}
	\end{center}
	\caption{Schematic of the method to construct the bulk or boundary state starting from an initial state which is a product state of quantum harmonic oscillator modes perpared in local thermal or ground (vacuum) states. The unitary operator $\hat{R}_{\rm bk}$ or $\hat{R}_{\rm bd}$ performs a sequence of passive linear transformations on mode operators for the bulk or boundary state construction. The unitary $\hat{D}$ is the same for both state constructions and involves a set of parallel single mode squeezing operations. If preparing bulk or boundary ground states, then the initial state is chosen to be the vacuum and the first round of linear operators $R^{\dagger}$ is not needed. 
}
	\label{fig:FigswithTransformMatrices}
\end{figure}

\section{Conclusions}
In summary, wavelets are a natural basis in which to study holographic duality in QFT. By way of the exact holographic mapping we have shown that ground states of $1+1$ dimensional scalar bosonic QFT can be represented in a wavelet basis where each DOF carries space-time and scale co\"ordinates. Correlations between the wavelet fields fall off exponentially with geodesic length where the metric is AdS$_3$ in the case of a massless boundary CFT and flat in the massive case. The boundary and bulk states arise as ground states of two body boundary or bulk Hamiltonians but they can also be constructed via a quantum circuit acting on a lattice of bosonic modes. The gate operations are Gaussian and the necessary amount of single mode squeezing grows logarithmically with system size meaning a proof of principle quantum simulation could be realized with engineered bosonic lattices. While properties of the bulk corresponding to boundary thermal states was not studied in this work such states are just as easily constructed by a quantum circuit.

There are several areas worthy of further investigation. First, we have only analysed the EHM for a single CFT with central charge $c=1$, which is the same central charge for the free Dirac fermion studied in \cite{Qi:13, QiLee:16}. In the pioneering work of Brown and Henneaux it was shown that gravitational theories in AdS$_3$ space of radius $R$ are dual to $1+1$ dimensional CFTs with central charge $c=3R/2G^{(3)}_N$ where $G^{(3)}_N$ is the Newton's gravitational constant in three spacetime dimensions. While we make no claim that the bulk description described here is a theory of gravity in $2+1$ dimensions, it would be worthwhile checking if there is a dependence of radius of curvature on central charge of the boundary CFT. Second, can the EHM provide useful insight for boundary gauge theories or interacting QFTs? It is possible to incorporate gauge fields with a wavelet decomposition as described in Refs. \cite{BP:13, AK2013} despite the fact that the basis functions are not strictly local and in Ref. \cite{BRSS:15} it was shown how interacting scalar bosonic QFT can be encoded in a wavelet basis and efficiently prepared and evolved on a quantum computer. Since the wavelet basis describes the state at multiple scales, this could be a useful means to infer properties of renormalization flow. Finally, can the wavelet description provide a better basis for tensor network descriptions of many body states or field theories? Recently, the wavelet transformation was used to construct the first \textit{analytic} MERA \cite{GlenWhite} for a critical system, meaning components of the constituent tensors were analytically derived from properties of the boundary theory (and not obtained e.g. from a numerical optimization). On the other hand, it is likely that a wavelet basis independent description of the bulk geometry could be derived by viewing any wavelet family as constructed from a $\emph{universal}$ nearest neighbor circuit like the binary MERA circuit. Indeed, entirely new classes of wavelets can be constructed using MERA like quantum circuits \cite{EW:16b}.  Further investigations may reveal other interesting connections between the wavelet and tensor network descriptions (e.g., see Ref.~\onlinecite{cMERA}) of quantum field theories.

\acknowledgements
We thank Barry Sanders and Glen Evenbly for helpful comments and feedback, and acknowledge many beneficial discussions with Dean Southwood.  We acknowledge support from the ARC via the Centre
of Excellence in Engineered Quantum Systems (EQuS),
project number CE110001013 and from DP160102426.
\appendix
\section{Boundary correlations}
\label{diagHambound}
\begin{figure}[t]
	\begin{center}
		\includegraphics[width=\columnwidth]{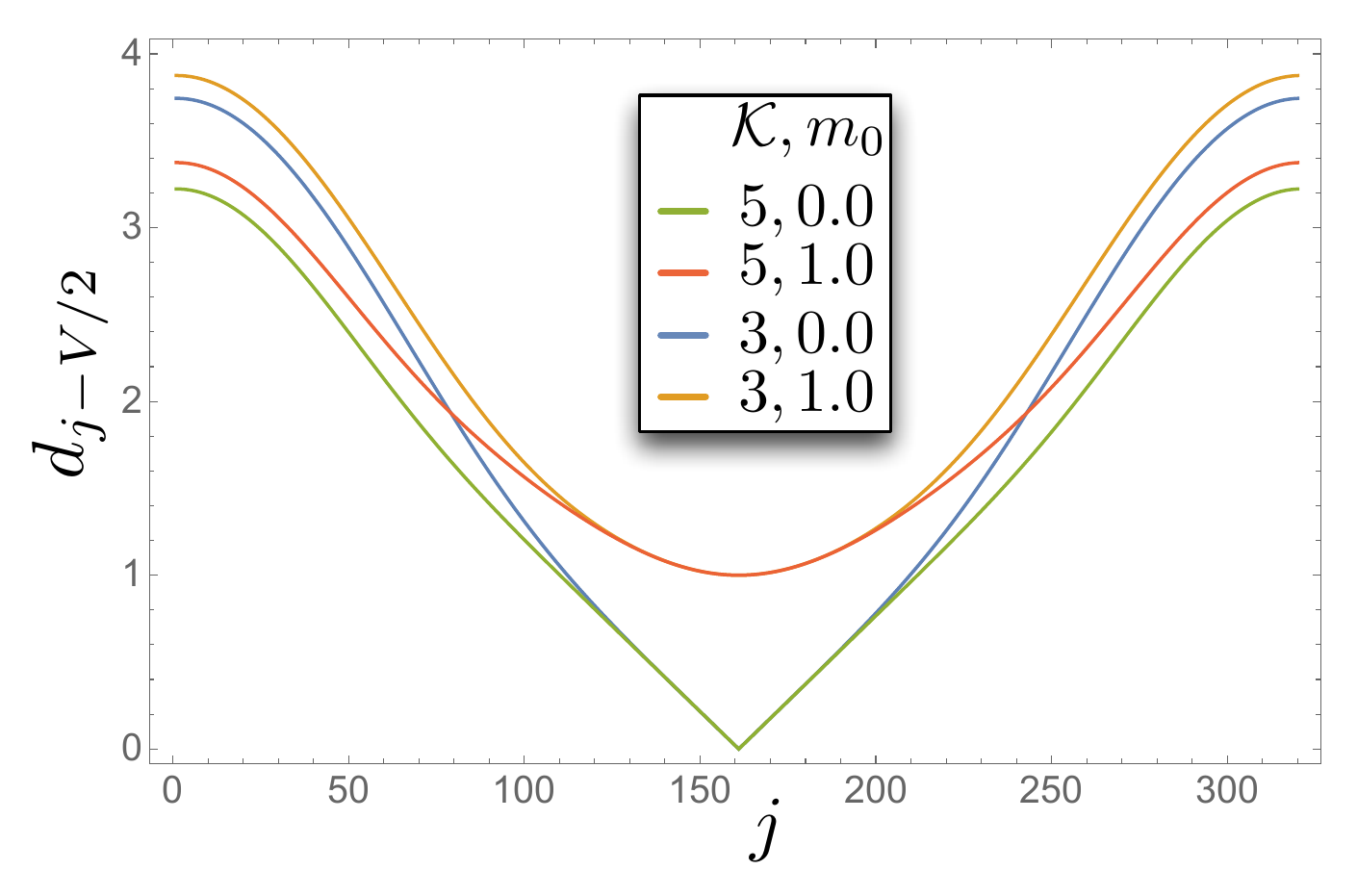}
	\end{center}
	\caption{Plot of the eigenvalues $d_j$ as a fucntion of $j$ (recall the momentum $k_j=\frac{2\pi j}{V}$), for a system of size $V=320$ and mass $m_0$ showing the dispersion relation obtained using Daubechies $\mathcal{K}=3,5$ wavelets. Note that larger $\mathcal{K}$ gives a better approximation to the linear dispersion obtained from the continuum model in the massless phase. 
}
	\label{fig:Dispersion}
\end{figure}

In this section we obtain the correlations for the scalar bosonic QFT at scale $n$ acting on $V=L\times 2^n$ modes.
The $V\times V$ matrix encoding the couplings in the boundary Hamiltonian can be written
\begin{equation}
K_{\rm bd}=(-D^{[ss]0}_{0,0} +m_0^2) {\bf 1}_V+\sum_{m=0}^{2\mathcal{K}-2}D^{[ss]0}_{0,m} (X_V^m+X_V^{\dagger m}), 
\end{equation}
where $X_V=\sum_{j=0}^{V-1} \ket{j\oplus_V 1}\bra{j}$ is the unitary increment operator.
The matrix can be diagonalized by a Fourier transform via the operator $F_V=\frac{1}{\sqrt{V}}\sum_{j,k=0}^{V-1} e^{i2\pi jk/V}\ket{j}\bra{k}$:
\[
K_{\rm bd} =F_V^{\dagger} [\oplus_{j=0}^{V-1}d^2_j] F_V,
\]
where the eigenvalues are the squares of
\[
d_j=\Big[(-D^{[ss]0}_{0,0} +m_0^2)+2\sum_{m=0}^{2\mathcal{K}-2}D^{[ss]0}_{0,m}\cos(k_j m)\Big]^{1/2},
\]
where we defined $k_j=\frac{2\pi j}{V}$ for $j\in \mathbb{Z}$ (see Fig. \ref{fig:Dispersion}).
Note in the massless case, the matrix is singular since $d_0=-D^{[ss]0}_{0,0}+2\sum_m  D^{[ss]0}_{0,m}=0$.
The boundary Hamiltonian can then be written in terms of normal modes
\[
\hat{H}_{\rm bd}=\sum_{j=0}^{V-1} d_j \left(\hat{\tilde{a}}^{[s]n\dagger}(j)\hat{\tilde{a}}^{[s]n}(j)+\frac{1}{2}\right),
\]
with normal mode operators $\hat{\tilde{a}}^{[s]n}(j)=\frac{1}{\sqrt{V}}\sum_{m=0}^{V-1}e^{i 2 \pi jm/V}\hat{a}^{[s]n}(m)$. In the massless
case the zero mode carries the annihiltion operator $\hat{\tilde{a}}^{[s]n}(0)$.
We define the ground state to be the unique state that satisfies $\hat{\tilde{a}}^{[s]n}(j)\ket{G}_{\rm bd}=0, \forall j$.

The correlations in the ground state are
\begin{equation}
\begin{array}{lll}
_{\rm bd}\bra{G}\hat{\Phi}^{\scale}(m)\hat{\Phi}^{\scale}(m')\ket{G}_{\rm bd}&=&\frac{1}{2}[K^{-1/2}_{\rm bd}]_{m,m'}\\
&=&\frac{1}{2V}\sum_{j=0}^{V-1} \frac{\cos(k_j (m-m') )}{d_j},
\end{array}
\label{bdFFcorrs}
\end{equation}
\begin{equation}
\begin{array}{lll}
_{\rm bd}\bra{G}\hat{\Pi}^{\scale}(m)\hat{\Pi}^{\scale}(m')\ket{G}_{\rm bd}&=&\frac{1}{2}[K^{1/2}_{\rm bd}]_{m,m'}\\
&=&\frac{1}{2V}\sum_{j=0}^{V-1} d_j\cos(k_j (m-m') ).
\end{array}
\label{bdMMcorrs}
\end{equation}
Numerically, we find that these correlations are well described by the continuum values in Eqs.~\ref{contcorrmassless},\ref{contcorrmassive} up to an additive constant that depends on mass and scale.

Similarly, the correlations in a thermal state $\hat{\rho}_{\rm bd}(\beta)$ at temperature $T=\beta^{-1}$ are
\begin{equation}
\tr[\hat{\Phi}^{\scale}(m)\hat{\Phi}^{\scale}(m')\hat{\rho}_{\rm bd}(\beta)]=\frac{1}{2V}\sum_{j=0}^{V-1} \frac{\coth(\beta d_j)}{d_j}\cos(k_j (m-m') ),
\end{equation}
\begin{equation}
\tr[\hat{\Pi}^{\scale}(m)\hat{\Pi}^{\scale}(m')\hat{\rho}_{\rm bd}(\beta)]=\frac{1}{2V}\sum_{j=0}^{V-1} d_j \coth(\beta d_j)\cos(k_j (m-m') ).
\end{equation}

For large $V$ the correlations on the boundary in the massless case for $|m-m'|>2\mathcal{K}-1$ are: 
\begin{equation}
\begin{array}{lll}
{_{\rm bd}\bra{G}}\hat{\Phi}^{\scale}(m)\hat{\Phi}^{\scale}(m')\ket{G}_{\rm bd}
&=& -\frac{\ln((m-m')^2)}{4\pi}+\kappa,\\
{_{\rm bd}\bra{G}}\hat{\Pi}^{\scale}(m)\hat{\Pi}^{\scale}(m')\ket{G}_{\rm bd}&=& 
-\frac{1}{2\pi (m-m')^2}.
\label{bdcorr}
\end{array}
\end{equation}
where $\kappa$ is a constant that depends on scale $n$ and $\mathcal{K}$. We will show below that this constant does not impact the relevant properties of the bulk. 
In the massive phase,
\begin{equation}
\begin{split}
_{\rm bd}\bra{G}\hat{\Phi}^{\scale}(m)\hat{\Phi}^{\scale}(m')\ket{G}_{\rm bd}&\approx \frac{1}{2\pi}K_0(m_0 |x-y|),\\
_{\rm bd}\bra{G}\hat{\Pi}^{\scale}(m)\hat{\Pi}^{\scale}(m')\ket{G}_{\rm bd}&\approx -\frac{m_0}{2\pi(x-y)}K_1(m_0 |x-y|),
\end{split}
\end{equation}

Temporal correlations are given by the boundary Green's function (time $\tau>0$):
\[
\begin{array}{lll}
&&{_{\rm bd}\bra{G}}\hat{a}^{[s]n}(m,\tau)\hat{a}^{[s]n\dagger}(m',0)\ket{G}_{\rm bd}=\frac{1}{V}\sum_{r,s=0}^{V-1} e^{i 2 \pi (m'r-ms)/V}\\
&&\quad\times {_{\rm bd}\bra{G}}\hat{\tilde{a}}^{[s]n}(s,\tau)\hat{\tilde{a}}^{[s]n\dagger}(r,0)\ket{G}_{\rm bd}\\
&&\quad=\frac{1}{V}\sum_{r,s=0}^{V-1} e^{i (2 \pi (m'r-ms)/V+d_s\tau)}{_{\rm bd}\bra{G}}\hat{\tilde{a}}^{[s]n}(s,0)\hat{\tilde{a}}^{[s]n\dagger}(r,0)\ket{G}_{\rm bd}\\
&&\quad =\frac{1}{V}\sum_{r,s=0}^{V-1} e^{i (2 \pi (m'r-ms)/V+d_s\tau)}\delta_{s,r}\\
&&\quad =\frac{1}{V}\sum_{r=0}^{V-1} e^{i (2 \pi (m'-m)r/V+d_r\tau)}.
\end{array}
\]
At this point we make a Wick rotation to imaginary time $\tau\rightarrow i\tau$.
In the massless case, and in the continuum limit where $d_j=2\pi j/V$, and with $\tau\gg 1$ then 
\[
{_{\rm bd}\bra{G}}\hat{a}^{[s]n}(m,\tau)\hat{a}^{[s]n\dagger}(m',0)\ket{G}_{\rm bd}\rightarrow\frac{\tau}{(m-m')^2+\tau^2}.
\]
In the wavelet basis, we find numerically that the correlations are
\begin{equation}
{_{\rm bd}\bra{G}}\hat{a}^{[s]n}(m,\tau)\hat{a}^{[s]n\dagger}(m',0)\ket{G}_{\rm bd}\approx \frac{0.32\tau}{(m-m')^2+\tau^2}.
\label{boundtempcorr}
\end{equation}

 \section{Bulk correlations}
 \subsection{Massless case}
\label{Bkcorrm0}
\subsubsection{Same scale correlations}
\label{samescalecorrs}
Using the bulk/boundary correspondence embodied in Eq.~\ref{corrs}, we can compute the bulk correlations from the boundary correlations. For the massless case, the same scale field/field correlation is (using \ref{bdcorr})
\begin{equation}
\begin{array}{lll}
C^{\hat{\Phi},\hat{\Phi}}((r,0),(r,j))
&=&2^{r-n}\displaystyle{\sum_{m=0}^{(2\mathcal{K}-1)\times 2^{n-r}}}\sum_{m'=j\times 2^{n-r}}^{(j+2\mathcal{K}-1)\times 2^{n-r}}\\
&&w_0^{0}((m+1)\times 2^{r-n})\\
&&w_0^{0}((m'-j\times 2^{n-r}+1)\times 2^{r-n})\\
&&\Big((1-\delta_{m,m'})(-\frac{2^{-n}\ln((m-m')^2)}{4\pi}+\kappa)\\
&&+\delta_{m,m'}C\Big).\\
\end{array}
\end{equation} 
Note, we have introduced a constant $C$ to make the correlations finite at zero separation on the boundary, the final result will be independent of $C$. This equation assumes $n-r\gg 0$ so let's write $m2^{r-n}\rightarrow x$, and treat $x$ as a continuous variable so that $\delta m= 2^{n-r}dx$ and sums are replaced by integrals: $\sum_{m=0}^{(2\mathcal{K}-1)\times 2^{n-r}}\rightarrow 2^{n-r}\int_0^{2\mathcal{K}-1} dx$. Then we find
\begin{equation}
\begin{array}{lll}
C^{\hat{\Phi},\hat{\Phi}}((r,0),(r,j))
&=&2^{n-r}\int_0^{2\mathcal{K}-1} dx \int_{j}^{j+2\mathcal{K}-1} dx' w_0^{0}(x) \\
&&w_0^{0}(x'-j)\Big((1-\delta(x-x'))\\
&&\times (-\frac{\ln((2^{2(n-r)}(x-x')^2)}{4\pi}+\kappa)\\
&&+\delta(x-x')C\Big).\\
\end{array}
\end{equation} 
Now assume $j>2\mathcal{K}-1$, i.e. focus on correlations longer range than the size of the wavelet modes, so that the integrals satisfy $\int_0^{2\mathcal{K}-1} dx w_0^{0}(x) w_0^{0}(x-j)=0$, and define $x''=x'-j$, then
\begin{equation}
\begin{array}{lll}
C^{\hat{\Phi},\hat{\Phi}}((r,0),(r,j))
&=&2^{n-r}\int dx \int dx'' w_0^{0}(x) w_0^{0}(x'')\\
&&\times \Big[(\kappa_n-\frac{2(n-r)\ln 2}{4\pi})-\frac{\ln((x-x''-j)^2)}{4\pi}\Big]\\
&=&-\frac{2^{n-r}}{4\pi}\int dx \int dx'' w_0^{0}(x) w_0^{0}(x'')\\
&&\times \ln((j-(x-x''))^2)\\
&=&-\frac{2^{n-r}}{2\pi}\int dx \int dx'' w_0^{0}(x) w_0^{0}(x'')\\
&&\times (\ln(1-(\frac{x-x''}{j}))+\ln (j))\\
&=&\frac{2^{n-r}}{2\pi}\int dx \int dx'' w_0^{0}(x) w_0^{0}(x'')\\
&&\times \sum_{k=1}^{\infty} \frac{(x-x'')^k}{k j^{k}}\\
&=&\frac{2^{n-r}}{2\pi}\sum_{k=1}^{\infty} \frac{1}{k j^{k}}\sum_{t=0}^k {k \choose t}(-1)^{t}\langle x^t\rangle_w \langle x^{k-t}\rangle_w.
\end{array}
\label{corrcomp}
\end{equation} 
Here we have defined the wavelet moments
\[
\langle x^a\rangle_w=\int x^a w^0_0(x)dx .
\] 
A method to calculate these moments is provided in Appendix \ref{momentscalc}.

A similar calculation gives the momenta-momenta correlations:
\begin{equation}
\begin{array}{lll}
C^{\hat{\Pi},\hat{\Pi}}((r,0),(r,j))
&=&-2^{n-r}\int dx \int dx'' w_0^{0}(x) w_0^{0}(x'')\\
&&\times \frac{1}{2\pi 2^{2(n-r)} (j-(x-x''))^2}\\
&=&-\frac{2^{r-n}}{2\pi j^2}\int dx \int dx'' w_0^{0}(x) w_0^{0}(x'')\\
&&\times \frac{1}{(1-(x-x'')/j)^2}\\
&=&-\frac{2^{r-n}}{2\pi j^2}\int dx \int dx'' w_0^{0}(x) w_0^{0}(x'')\\
&&\times \frac{1}{(1-(x-x'')/j)^2}\\
&=&-\frac{2^{r-n}}{2\pi j^2}\int dx \int dx'' w_0^{0}(x) w_0^{0}(x'')\\
&&\sum_{k=0}^{\infty} \frac{k+1}{j^k} (x-x'')^k\\
&=&-\frac{2^{r-n}}{2\pi j^2}\sum_{k=0}^{\infty} \frac{k+1}{j^k} \\
&&\sum_{t=0}^k {k \choose t}(-1)^{t}\langle x^t\rangle_w \langle x^{k-t}\rangle_w.
\end{array}
\label{mommomcorr}
\end{equation} 
Keeping only the dominant term, for $j>2\mathcal{K}-1$, 
\begin{equation}
C^{\hat{\Phi},\hat{\Phi}}((r,0),(r,j))\approx -\frac{2^{n-r} \times D_{\mathcal{K}}}{4\pi \mathcal{K}j^{2\mathcal{K}}},
\label{phiphicorrapp}
\end{equation}
where
\begin{equation}
D_{\mathcal{K}}=\langle x^{\mathcal{K}}\rangle_w^2{2\mathcal{K} \choose \mathcal{K}}.
\end{equation}
Numerically we find the self correlation
\begin{equation}
C^{\hat{\Phi},\hat{\Phi}}((r,0),(r,0))= 2^{n-r-a}.
\label{pipicorrapp}
\end{equation}
Similarly, for the momentum-momentum correlations, for $j>2\mathcal{K}-1$,
\begin{equation}
C^{\hat{\Pi},\hat{\Pi}}((r,0),(r,j))\approx \frac{2^{r-n}\times (2\mathcal{K}+1)D_{\mathcal{K}}}{2\pi j^{2\mathcal{K}+2}}.
\end{equation}
The self correlations can be written
\begin{equation}
C^{\hat{\Pi},\hat{\Pi}}((r,0),(r,0))= 2^{r-n+b}.
\end{equation}
The values $a,b$ appearing in the self correlations can be calculated using Eq.~\ref{corrs} together with Eqs.~\ref{bdFFcorrs},\ref{bdMMcorrs}. By computing self correlations for $ 3\leq  \mathcal{K}\leq 30$ we find that $a\approx 3.18$ independent of $\mathcal{K}$, whereas $b$ is fit by the function $b\approx 0.75/\mathcal{K}^2+0.16 \mathcal{K}+1.24$. 
In order for the reduced covariance matrix of a single bulk site to describe a valid quantum state, it should satisfy the positivity condition
\[
\Gamma_1+\frac{i}{2}\Omega\geq 0,
\]
where $\Omega$ is the symplectic form from Eq.~\ref{BigOmega} (with $n_A=1$) and 
\[
\Gamma_1=
\left(\begin{array}{cc} C^{\hat{\Phi},\hat{\Phi}}((r,0),(r,0)) & 0 \\0 & C^{\hat{\Pi},\hat{\Pi}}((r,0),(r,0)) \end{array}\right).
\]
It is readily verified that the computed functional form for the self correlations does satisfy positivity.

\subsubsection{Temporal correlations}
\label{apptempcorrs}
For simplicity, we consider temporal correlations at the same spatial point $(r,0)$. Following the same argument in Sec. \ref{samescalecorrs}, Then following the same argument as that leading to Eq.~\ref{corrs} we find deep in the bulk
\begin{equation}
\begin{array}{lll}
C^{\hat{a},\hat{a}^{\dagger}}((r,0,\tau),(r,0,0))
&=&2^{r-n}\displaystyle{\sum_{m=0}^{(2\mathcal{K}-1)\times 2^{n-r}}}\sum_{m'=0}^{(2\mathcal{K}-1)\times 2^{n-r}}\\
&&w_0^{0}((m+1)\times 2^{r-n}) \\
&&w_0^{0}((m'+1)\times 2^{r-n})\\
&&{_{\rm bd}\bra{G}}T[\hat{a}^{\scale}(m,\tau)\hat{a}^{\dagger\scale}(m',0)]\ket{G}_{\rm bd}.\\
\end{array}
\label{tempcorrs}
\end{equation}
Using the boundary Green's function from Eq.~\ref{boundtempcorr}, and assuming we are deep in the bulk so that we can approximate the sum by an integral as in Sec. \ref{samescalecorrs}, and defining $\tau 2^{r-n}=\tilde{\tau}$ then
\begin{equation}
\begin{array}{lll}
C^{\hat{a},\hat{a}^{\dagger}}((r,0,\tau),(r,0,0))
&=&0.32 \tilde{\tau}\int_0^{2\mathcal{K}-1} dx \int_{0}^{2\mathcal{K}-1} dx' w_0^{0}(x) \\
&&w_0^{0}(x')((x-x')^2+\tilde{\tau}^2)^{-1}\\
&=&0.32\tilde{\tau}^{-1}\int_0^{2\mathcal{K}-1} dx \int_{0}^{2\mathcal{K}-1} dx'w_0^{0}(x)\\
&&w_0^{0}(x')\sum_{s=0}^{\infty}(-1)^s(\frac{x-x'}{\tilde{\tau}})^{2s}\\
&=&0.32\tilde{\tau}^{-1}\sum_{s=0}^{\infty}\sum_{k=0}^{2s} (-1)^{s+k} \tilde{\tau}^{-2s}\\
&&{2s \choose k}\langle x^k\rangle_w \langle x^{2s-k}\rangle_w.
\end{array}
\end{equation}
In the second line we have assumed that $\tau>2^{n-r}(2\mathcal{K}-1)$ in order to perform the power series expansion. Keeping only the dominant term,
\begin{equation}
C^{\hat{a},\hat{a}^{\dagger}}((r,0,\tau),(r,0,0))\approx \frac{0.32\times D_{\mathcal{K}}}{(2^{r-n}\tau)^{(2\mathcal{K}+1)}}.
\end{equation}

\subsection{Massive case}
\label{massive}
In the massive phase, the boundary ground state satisfies 
\begin{equation}
\begin{split}
_{\rm bd}\bra{G}\hat{\Phi}^{\scale}(m)\hat{\Phi}^{\scale}(m')\ket{G}_{\rm bd}&=\frac{1}{2\pi}K_0(m_0 |x-y|),\\
_{\rm bd}\bra{G}\hat{\Pi}^{\scale}(m)\hat{\Pi}^{\scale}(m')\ket{G}_{\rm bd}&= -\frac{m_0}{2\pi(x-y)}K_1(m_0 |x-y|),
\end{split}
\end{equation}
for $m=(0,1,\ldots L2^n/2-1)$. As in Sec. \ref{RadiusComp} we can compute the same scale bulk correlations $C^{\hat{\Phi},\hat{\Phi}}((r,0),(r,j))$. Following the same argument that led to Eq.~\ref{corrcomp}, and again assuming that $j>2\mathcal{K}-1$, we find for the same scale correlations
\begin{equation}
\begin{array}{lll}
C^{\hat{\Phi},\hat{\Phi}}((r,0),(r,j))
&=&\frac{2^{n-r}}{2\pi}\int dx \int dx'' w_0^{0}(x) w_0^{0}(x'')\\
&&\times K_0(m_0 2^{n-r} (j-(x-x'')).
\end{array}
\end{equation} 
For $m_02^{n-r}\gg 1$, this simplifies to 
\begin{equation}
\begin{array}{lll}
C^{\hat{\Phi},\hat{\Phi}}((r,0),(r,j))
&=&-\frac{2^{n-r}e^{-j m_0 2^{n-r}}}{\sqrt{8\pi 2^{n-r} m_0}}\int dx \int dx'' w_0^{0}(x) w_0^{0}(x'')\\
&&\times \frac{e^{-m_0 2^{n-r} (x''-x)}}{\sqrt{j-(x-x'')}}\\
&\approx &-\frac{2^{n-r}e^{-j m_0 2^{n-r}}}{\sqrt{j 8\pi 2^{n-r} m_0}}\int dx \int dx'' w_0^{0}(x) w_0^{0}(x'')\\
&&\times e^{-m_0 2^{n-r} (x''-x)}\\
&=&-\frac{2^{n-r}e^{-j \tilde{m}}}{\sqrt{j 8\pi \tilde{m}}}\langle e^{-\tilde{m} x}\rangle_w \langle e^{\tilde{m}x}\rangle_w.\\
\end{array}
\end{equation} 
In the second line we have assumed that $j\gg 2\mathcal{K}-1$ in order to simplify the denominator.  The error introduced in doing so is small when $m_02^{n-r}\gg 1$.  
The same scale bulk correlations fall off exponentially with a renormalized mass
\[
\tilde{m}=m_0 2^{n-r}.
\]
The momentum-momentum correlations on the other hand, in the same limit $\tilde{m}\gg 1$, are:
\begin{equation}
\begin{array}{lll}
C^{\hat{\Pi},\hat{\Pi}}((r,0),(r,j))
&=&\frac{2^{n-r}e^{-j m_0 2^{n-r}}\sqrt{m_0}}{\sqrt{8\pi 2^{3(n-r)}}}\int dx \int dx'' w_0^{0}(x) w_0^{0}(x'')\\
&&\times \frac{e^{-m_0 2^{n-r} (x''-x)}}{\sqrt{(j-(x-x''))^3}}\\
&\approx &2^{r-n}e^{-j \tilde{m}}\sqrt{\frac{\tilde{m}}{8\pi j^3}}\langle e^{-\tilde{m} x}\rangle_w \langle e^{\tilde{m} x}\rangle_w.
\end{array}
\end{equation} 
For $\mathcal{K}=3$ wavelets and $\tilde{m}\gg 1$ the product of averages is $\langle e^{-\tilde{m} x}\rangle_w \langle e^{\tilde{m} x}\rangle_w\approx e^{4.71 \times \tilde{m}-20.38}$.
\section{Distances in AdS$_{3}$ space}
\label{AdSDistance}
The metric of Euclidean AdS$_{3}$ space (with time co\" ordinate $t\rightarrow i\tau$) is
\begin{equation}
ds^2=\left(\frac{\rho^2}{R^2}+1\right)d\tau^2+\frac{1}{\frac{\rho^2}{R^2}+1}d\rho^2+\rho^2d\theta^2,
\end{equation}
where $\rho\in\mathbb{R}^+$ is a radial co\"ordinate, $\theta\in[0,2\pi)$ is the angular co\"ordinate, $\tau\in\mathbb{R}$, and $R$ is the radius of curvature.
On a time slice, the geodesic distance between two points at the same radius $\rho$ is:
\begin{equation}
\begin{array}{lll}
d_g((\rho,\theta_1),(\rho,\theta_2))&=&R\cosh^{-1}\left(1+\frac{2\rho^2}{R^2}\sin^2(\frac{\theta_1-\theta_2}{2})\right)\\
&=&R\ln\Big[1+\frac{2\rho^2}{R^2}\sin^2(\frac{\theta_1-\theta_2}{2})\\
&&+\sqrt{(1+\frac{2\rho^2}{R^2}\sin^2(\frac{\theta_1-\theta_2}{2}))^2-1}\Big]\\
&\approx&R\ln[\frac{4\rho^2}{R^2}\sin^2(\frac{\theta_1-\theta_2}{2})]\\
&=&2R\ln[\frac{2\rho}{R}\sin(\frac{\theta_1-\theta_2}{2})]\\
&\approx&2R\ln[\frac{\rho(\theta_1-\theta_2)}{R}]
\label{geodesicangular}
\end{array}
\end{equation}
where in third line we have assumed that we are deep in the bulk and with sufficiently small radius of curvature so that $\rho\gg R$ and in the fifth line we have assumed we are considering angular separations $|\theta_1-\theta_2|\ll \pi$. In our bulk description, fields localized at scale $r$ and position $m$ have AdS$_{3}$ co\"ordinates $(\rho=\frac{L2^{r}}{2\pi},\theta=\frac{m2\pi}{L2^{r}})$, so the geodesic distance between these co\"ordinates is
\begin{equation}
 d_g((r,m),(r,m'))\approx 2R\ln\left[\frac{|m-m'|}{R}\right],
 \label{geosamescale}
 \end{equation}
where $R$ has the same dimensions as position $m$ (we make them dimensionless). To find the geodesic distance between points at the same angle but different radii, first find the distance to the origin from a point at $(\rho,\theta)$ which is one half the distance computing using Eq.~\ref{geodesicangular} with antipodal points $\theta_1-\theta_2=\pi$. This gives $d_g((r,0),(0,0))\approx R\left(\ln[\frac{L}{2\pi R}]+r\ln 2\right)$. Then the geodesic distance between points is
\begin{equation}
d_g((r,0),(r',0))\approx R|r'-r| \ln 2 .
\label{geoacrossscales}
\end{equation}
where recall scale $r$ is dimensionless.

Finally, the geodesic distance can be computed between two events with the same spatial co\"ordinates but separated in imaginary time by $\tau$. The time co\"ordinate in Euclidean AdS$_{3}$ must be rescaled according to $\tau\rightarrow \tau/\sqrt{R^2+V/((2\pi))^2}$ where $V$ is the size of the boundary, in order that the metric reduces to $ds^2=d\tau^2+\rho^2d\theta^2$ at the boundary $(\rho=V/2\pi)$. When this is done then for $\rho\gg R$, $\tau\ll V$, and $\rho \tau\gg RV$,
\begin{equation}
d_g((\rho,\theta,\tau),(\rho,\theta,0)\approx 2R\ln\left(\frac{2\pi \rho \tau}{RV}\right).
\end{equation}
In terms of the bulk geometry co\"ordinates with boundary size $V=L2^n$ then 
\begin{equation}
d_g((r,j,\tau),(r,j,0))\approx 2R\ln\left(\frac{\tau 2^{n-r}}{R}\right).
\end{equation}

\section{Properties of Daubechies wavelets}
\label{waveletprops}
The family of Daubechies wavelets are constructed from a set of scale function coefficients $\{h_j\}_{j=0}^{2\mathcal{K}-1}$, and
wavelet coefficients $\{g_j=(-1)^j h_{2\mathcal{K}-1-j}\}$.
The coefficients for many families are available in the literature, e.g. \cite{Daubechies}, and in several software packages. We provide them for three families in Table \ref{table:scalefunctioncoeffs}, and plots of the scale and wavelet functions for $\mathcal{K}=3,4,5$ are shown in Fig. \ref{fig:WaveletPlots}.
\subsection{Derivative overlaps}
\label{deroverlap}
The procedure to compute the derivative overlaps in Eq.~\ref{deriviativeoverlaps} is given in Ref. \cite{BP:13} and for completeness we
include it here and evaluate them for several wavelet families.
\begin{table}[H]
\centering 
\begin{tabular}{c | c | c | c } 
\hline\hline 
$h_j$ & $\mathcal{K}=3$ & $\mathcal{K}=4$ & $\mathcal{K}=5$ \\ [0.5ex] 
\hline
$h_0$ & $\frac{1+\sqrt{10}+\sqrt{5+2\sqrt{10}}}{16\sqrt{2}}$  & 0.230377813308897 & 0.160102397974193 \\
$h_1$ & $\frac{5+\sqrt{10}+3\sqrt{5+2\sqrt{10}}}{16\sqrt{2}}$  & 0.714846570552916 & 0.603829269797190 \\
$h_2$ & $\frac{10-2\sqrt{10}+2\sqrt{5+2\sqrt{10}}}{16\sqrt{2}}$  & 0.630880767929859 & 0.724308528437773 \\
$h_3$ & $\frac{1+\sqrt{10}+\sqrt{5+2\sqrt{10}}}{16\sqrt{2}}$  & -0.027983769416860 & 0.138428145901321 \\
$h_4$ & $\frac{5+\sqrt{10}-3\sqrt{5+2\sqrt{10}}}{16\sqrt{2}}$  & -0.187034811719093 & -0.242294887066382 \\
$h_5$ & $\frac{1+\sqrt{10}-\sqrt{5+2\sqrt{10}}}{16\sqrt{2}}$  & 0.030841381835561 & -0.032244869584638 \\
$h_6$ & 0  & 0.032883011666885 & 0.077571493840046 \\
$h_7$ & 0  & -0.010597401785069 & -0.006241490212798 \\
$h_8$ & 0  & 0 & -0.012580751999082 \\
$h_9$ & 0  & 0 & 0.003335725285474\\
[1ex]
\hline 
\end{tabular}
\caption{
Values of the scale function coefficients for the wavelet families $\mathcal{K}=3,4,5$.
\label{table:scalefunctioncoeffs} 
}
\end{table}

The overlap between derivatives of scale/scale functions is given by the coefficients $D^{[ss]k}_{m,m'}$ which satisfy $D^{[ss]k}_{m,m'}=2^{2k}D^{[ss]0}_{0,m'-m}$ so it suffices to find the coeffiecients $D^{[ss]0}_{0,m}$. The cofficients are symmetric $D^{[ss]0}_{0,m}=D^{[ss]0}_{m,0}$ and because of the compact support of the scale functions, are non zero only for $|m|\leq 2\mathcal{K}-2$. Using a resolution of the identity $\sum_n s^0_n(x)=1$, the coefficients can be written
\begin{equation}
D^{[ss]0}_{0,m}=\sum_n D_{n,0,m}=\sum_n D_{0,-n,m-n},
\end{equation}
where $D_{n,j,k}=\int dx s^0_n(x)\partial_x s^{0}_{j}(x)\cdot \partial_x s^{0}_{k}(x)$. These coefficients are symmetric in the last two indices $D_{n,j,k}=D_{n,k,j}$ and satisfy the following set of homogeneous equations
\begin{equation}
\begin{array}{lll}
\left\{D_{0,r,s}=4\sqrt{2}\sum_{n=0}^{2\mathcal{K}-1}\sum_{j,k=-(2\mathcal{K}-2)}^{2\mathcal{K}-2}h_{n}h_{k+n-2r}h_{j+n-2s}D_{0,k,j}\right\}\\
\cup \left\{\sum_{k=-(2\mathcal{K}-2)}^{2\mathcal{K}-2}D_{0,j,k}=0\right\};\ r,s\in[-(2\mathcal{K}-2),2\mathcal{K}-2]\cap\mathbb{Z},
\end{array}
\end{equation}
and the set of inhomogeneous equations
\begin{equation}
\left\{\sum_{j=-(2\mathcal{K}-2)}^{2\mathcal{K}-2}jD_{0,j,k}=\Gamma_{0,k}\right\};\ k\in[-(2\mathcal{K}-2),2\mathcal{K}-2]\cap\mathbb{Z},
\end{equation}
where $\Gamma_{n,m}=\int dx s^0_n(x)\partial_x s^0_m(x)=-\Gamma_{m,n}=\Gamma_{0,m-n}$. The coefficients $\Gamma_{n,m}$ themselves satisfy the following set of homogeneous equations
\begin{equation}
\begin{split}
&\left\{\Gamma_{0,j}=2\sum_{m=0}^{2\mathcal{K}-1}\sum_{n=-(2\mathcal{K}-2)}^{2\mathcal{K}-2}h_{m}h_{n+m-2j}\Gamma_{0,n}\right\};\\
&j\in[-(2\mathcal{K}-2),2\mathcal{K}-2]\cap\mathbb{Z},
\end{split}
\end{equation}
and one inhomogeneous equation
\begin{equation}
\sum_{n=-(2\mathcal{K}-2)}^{2\mathcal{K}-2} n\Gamma_{0,n}=1.
\end{equation}
First solving for the $\Gamma_{0,m}$ one can then solve for the $D_{0,r,s}$ and finally for $D^{[ss]}_{0,m}$. Note, the coefficients satisfy $D^{[ss]}_{0,m}=D^{[ss]}_{0,-m}$, and also $\sum_{m=0}^{2\mathcal{K}-2} m^2 D^{[ss]0}_{0,m}=-1$.
The other overlaps of derivatives of scale/wavelet functions and wavelet/wavelet functions are given by \cite{BRSS:15}

\begin{equation}
\begin{array}{lll}
D^{[sw]l,0}_{a,b}&=&2^{2(l+1)}(\bra{a} [H(l)]^{l+1} D(l) G^T(l)\ket{b}\\
&&+\bra{a+L} [H(l)]^{l+1} D(l) G^T(l)\ket{b}\\
&&+\bra{a} [H(l)]^{l+1} D(l) G^T(l)\ket{b+2^l L}),
\\
D^{[ww]l,j}_{a,b}&=&2^{2(l+1)}(\bra{a} G(l,j)[H(l,j)]^{l-j} D(l,j) G^T(l,j)\ket{b}\\
&&+\bra{a+2^j L} G(l,j)[H(l,j)]^{l-j} D(l,j) G^T(l,j)\ket{b}\\
&&+\bra{a} G(l,j)[H(l,j)]^{l-j} D(l,j) G^T(l,j)\ket{b+2^l L}),
\end{array}
\end{equation}
where the scale dependent matrices are 
\begin{equation}
\begin{array}{lll}
H(l)&=&\sum_{m,n=0}^{2^{(l+2)}(L+2\mathcal{K}-2)-(2\mathcal{K}-1)}h_{n-2m}\ket{m}\bra{n} \\
H(l,j)&=&\sum_{m,n=0}^{2^{(l-j+1)}2 (2^j L-2\mathcal{K}-2))-(2\mathcal{K}-1)}h_{n-2m}\ket{m}\bra{n} \\
D(l)&=&\sum_{m,n=0}^{2^{(l+2)}(L+2\mathcal{K}-2)-(2\mathcal{K}-1)}D^{[ss]0}_{m,n}\ket{m}\bra{n} \\
D(l,j)&=&\sum_{m,n=0}^{2^{(l-j+1)}2 (2^j L-2\mathcal{K}-2)-(2\mathcal{K}-1)}D^{[ss]0}_{m,n}\ket{m}\bra{n} \\
G(l)&=&\sum_{m,n=0}^{2^{(l+2)}(L+2\mathcal{K}-2)-(2\mathcal{K}-1)}g_{n-2m}\ket{m}\bra{n} \\
G(l,j)&=&\sum_{m,n=0}^{2^{(l-j+1)}2 (2^j L-2\mathcal{K}-2)-(2\mathcal{K}-1)}g_{n-2m}\ket{m}\bra{n}. \\
\end{array}
\end{equation}
Note at different scales these coefficients satisfy $D^{[ww]r-n,r'-n}_{m,m'}=2^{-2n}D^{[ww]r,r'}_{m,m'}$, and $D^{[sw]r-n,-n}_{m,m'}=2^{-2n}D^{[sw]r,0}_{m,m'}$.

\subsection{Wavelet moments}
\label{momentscalc}
The wavelet moments
\[
\langle x^a\rangle_w=\int x^a w^0_0(x) dx,
\]
satisfy the relation \cite{BP:13}
\begin{equation}
\langle x^a\rangle_w=\frac{1}{\sqrt{2}}\frac{1}{2^a}\sum_{j=0}^{2\mathcal{K}-1} g_j \sum_{b=0}^a {a \choose b}j^{a-b}\langle x^b\rangle_s,
\end{equation}
where $g_j=(-1)^j h_{2\mathcal{K}-1-j}$, \footnote{Note we have defined $0^0=1$.} and $h_k$ are the aforementioned scale functions coefficients. 
The scale function moments 
\[
\langle x^b\rangle_s=\int x^b s^0_0(x)dx,
\]
satisfy a recursion formula
\begin{equation}
\langle x^b\rangle_s=\frac{1}{2^b-1}\frac{1}{\sqrt{2}}\sum_{c=0}^{b-1} {b \choose c}\sum_{k=1}^{2\mathcal{K}-1} h_k k^{b-c} \langle x^c\rangle_s,
\end{equation}
with $\langle x^0\rangle_s=1$. The $\mathcal{K}$ wavelets have vanishing moments up to $\mathcal{K}-1$.

For $\mathcal{K}=3$, the lowest few wavelet moments are
\[
\begin{array}{lll}
\langle x^a\rangle_w&=&0,\quad a=0,1,2\\
\langle x^3\rangle_w&=&-\frac{3\sqrt{5/2}}{16}\\
\langle x^4\rangle_w&=&\frac{3}{32}\Big(\sqrt{10}(\sqrt{5+2\sqrt{10}}-10)-\sqrt{5+2\sqrt{10}}\Big)\\
\langle x^5\rangle_w&=&\frac{15}{64}\Big(-3-26\sqrt{10}-5\sqrt{5+2\sqrt{10}}+5\sqrt{50+20\sqrt{10}}\Big)\\
\langle x^6\rangle_w&=&\frac{15}{896}\Big(-523\sqrt{5+2\sqrt{10}}+541\sqrt{50+20\sqrt{10}}\\
&&-70(9+28\sqrt{10})\Big).\\
\end{array}
\]

\begin{table}[H]
\centering 
\begin{tabular}{c | c | c | c } 
\hline\hline 
$D^{[ss]0}_{0,m}$ & $\mathcal{K}=3$ & $\mathcal{K}=4$ & $\mathcal{K}=5$ \\ [0.5ex] 
\hline
$D^{[ss]0}_{0,0}$ & 5.267857142856938 & 4.165973640640697 & 3.834994313804270 \\
$D^{[ss]0}_{0,1}$ & -3.390476190475967 & -2.642070208104849 & -2.414790351189328 \\
$D^{[ss]0}_{0,2}$ & 0.876190476190400  & 0.697869104358090 & 0.649502190049664 \\
$D^{[ss]0}_{0,3}$ & -0.114285714285703 & -0.150972899616039 & -0.180953550093395 \\
$D^{[ss]0}_{0,4}$ & -0.005357142857147  & 0.010572727778006 & 0.029907980478582 \\
$D^{[ss]0}_{0,5}$ & 0  & 0.001630376885712 & -0.000794620509733 \\
$D^{[ss]0}_{0,6}$ & 0  & -0.000015921649278 & -0.000367145399803 \\
$D^{[ss]0}_{0,7}$ & 0  & 0 & $-1.656544978 \times 10^{-6}$ \\
$D^{[ss]0}_{0,8}$ & 0  & 0 & $-3.574532\times 10^{-9}$ \\
[1ex]
\hline 
\end{tabular}
\caption{
Derivative overlap coefficients. Note: $D^{[ss]}_{0,m}=D^{[ss]}_{0,-m}$.
\label{table:overlapcoeffs} 
}
\end{table}

\end{document}